\documentclass[12pt,preprint]{aastex}

\usepackage{graphics,epsfig}
\usepackage{natbib}
\usepackage{lscape}

\def\spose#1{\hbox to 0pt{#1\hss}}
\def\lta{\mathrel{\spose{\lower 3pt\hbox{$\mathchar"218$}}
     \raise 2.0pt\hbox{$\mathchar"13C$}}}
\def\gta{\mathrel{\spose{\lower 3pt\hbox{$\mathchar"218$}}
     \raise 2.0pt\hbox{$\mathchar"13E$}}}

\def\figure#1#2 {\par{\narrower\noindent {\bf Fig. #1}
   \hskip 2mm #2\par}\bigskip\noindent}
\def\table#1#2 {\par{\narrower\noindent {\bf Tab. #1}
   \hskip 2mm #2\par}\bigskip\noindent}

\shorttitle{Habitability around F-type Stars}

\shortauthors{Sato et al.}

\begin{document}


\title{Habitability around F-type Stars}


\vspace{2.5cm}
\author{S. Sato$^1$, M. Cuntz$^1$, C. M. Guerra Olvera$^2$, D. Jack$^2$, 
K.-P. Schr\"oder$^2$}
\vspace{2.5cm}
\affil{$^1$Department of Physics, University of Texas at Arlington,
           Arlington, TX 76019, USA}
\vspace{0.5cm}
\affil{$^2$Department of Astronomy, University of Guanajuato, 
36000 Guanajuato, GTO, Mexico}
\vspace{0.5cm}
\email{satoko.sato@mavs.uta.edu; cuntz@uta.edu; cguerra@astro.ugto.mx; 
dennis@astro.ugto.mx; kps@astro.ugto.mx}


\clearpage


\begin{abstract}
We explore the general astrobiological significance of F-type main-sequence stars
with masses between 1.2 and 1.5~$M_\odot$.  Special consideration is given to
stellar evolutionary aspects due to nuclear main-sequence evolution.  DNA is
taken as a proxy for carbon-based macromolecules following the paradigm
that extraterrestrial biology may be most likely based on hydrocarbons.
Consequently, the DNA action spectrum is utilized to represent the impact
of the stellar UV radiation.  Planetary atmospheric attenuation is taken
into account based on parameterized attenuation functions.  We found that
the damage inflicted on DNA for planets at Earth-equivalent positions is
between a factor of 2.5 and 7.1 higher than for solar-like stars, and there
are intricate relations for the time-dependence of damage during stellar
main-sequence evolution.  If attenuation is considered, smaller factors of
damage are obtained in alignment to the attenuation parameters.  This work
is motivated by earlier studies indicating that the UV environment
of solar-type stars is one of the most decisive factors in determining the
suitability of exosolar planets and exomoons for biological evolution and
sustainability.
\end{abstract}


\keywords{
extra-solar planets, extraterrestrial life, F-type stars, habitable zone,
planetary climate and stellar evolution.
}


\clearpage

\section{Introduction}

Omitting O and B-type stars in consideration of their short lifespans
and strong ionizing winds \citep[e.g.,][]{maed88,kudr00}, as well as
the transitory case of A-type stars, F-type main-sequence stars represent
the hot limit of stars with a significant potential for providing
circumstellar habitable environments \citep[e.g.,][]{kast93,unde03}.
Generally, the investigation of habitability around different types of stars,
particularly main-sequence stars, is considered a theme of pivotal interest
to the thriving field of astrobiology; see, e.g., \cite{jone08}, \cite{lamm09},
\cite{kalt10}, \cite{horn10}, and \cite{lamm13} for previous studies and
reviews.  For planetary host stars --- like any other star --- the stellar
mass determines their lifetime (shorter with larger mass),
luminosity and effective temperature (both higher with larger mass).
Main-sequence stars moderately more massive than the Sun with masses between
1.1 and 1.6~$M_\odot$, i.e., F-type stars, are of particular interest as hosts
to exosolar planets and exomoons in orbit about those planets.  Compared to
stars of later spectral types, F-type stars are characterized by relatively
large habitable zones, although from a general astrobiological point of view,
they exhibit the adverse statistical property of being less frequent
\citep[e.g.,][]{krou02,chab03}.  On the other hand, despite their reduced
lifetimes compared to G-type stars, their lifetimes still exceed several
billion years \citep[e.g.,][]{maed88}, allowing the principle possibility
of exobiology, potentially including advanced life forms.

While habitable zones and their evolution can, in general, be well
characterized in terms of the total amount of stellar irradiation, the
spectral energy distribution (including its portion of energetic radiation)
may nevertheless be significant as well for the facilitation of habitability
\citep[e.g.,][]{lamm09}.
The emergent radiation of F-type stars consists of significantly
larger amounts of UV compared to the Sun thus entailing potentially unfavorable
effects on planetary climates and possible organisms \citep[e.g.,][]{cock99}.
Previous studies showed that increased levels of UV, as well as the even more
energetic EUV radiation, can trigger a variety of chemical planetary atmospheric
processes, including exoplanetary atmospheric evaporation
\citep[e.g.,][]{guin02,lamm03,gued07}.  Hence, the radiation output provided
by F-stars places an additional constraint on circumstellar habitability,
which needs to be considered as part of more comprehensive assessments;
see, e.g., \cite{bucc06} for previous results.
It is the aim of the present study to consider some of these processes in
an approximate manner while also taking into account the evolutionary status
of the F-type host stars.

In Sect.~2, we comment on the concept of habitability, which also includes
a description of the climatological habitable zone.  Additionally, we discuss
the governing equations of the present study involving both the DNA action
spectrum and planetary atmospheric attenuation.  In Sect.~3, we describe
the spectral energy  distribution of the host stars based on sophisticated
photospheric models and their spectral energy output computed by the PHOENIX
code for the range of effective temperatures relevant to F-type stars.
The specific effective temperature and total luminosity of each star for
a given mass and age are derived from well-tested stellar evolution models,
which also convey the time scales of the circumstellar conditions during
stellar main-sequence evolution as well as the extents of the climatological
habitable zones.  Results and discussion are given in Sect.~4, which focuses
on habitability gauged by the damage inflicted upon DNA.  Particular emphasis
is placed on the relevance of the different types of UV (i.e., UV-A, UV-B, and UV-C;
see Sect.~2.2 for definitions).  Our summary and conclusions are given in Sect.~5.
   

\section{Concepts of Habitability}

\subsection{Climatological Habitable Zone}

A key aspect in the study of circumstellar habitability is the introduction
of the climatological habitable zone, a concept, evaluated by \cite{kast93}.
They utilized 1-D climate model to estimate the position and width of
the habitable zone around solar-like stars as well as other types of
main-sequence stars.  The basic premise consists in assuming a Earth-like
planet with a CO$_2$/H$_2$O/N$_2$ atmosphere and, furthermore, that
habitability requires the presence of water on the planetary surface.
In their work they distinguished between the {\it conservative} habitable zone
(CHZ, with limits of 0.95 and 1.37~AU) and the {\it general} habitable zone 
(GHZ, with limits of 0.84 and 1.67~AU); subsequent work about the GHZ has also
been given by \cite{forg97} and others\footnote{Alternate usages of the acronym
CHZ include climatological habitable zone and continuous habitable zone.  Also,
an alternate usage of the acronym GHZ includes galactic habitable zone.  
However, those do not apply in this study.}.

The physical significance of the various kinds of HZs obtained
by \cite{kast93} are given as follows:  The GHZ is defined
as bordered by the runaway greenhouse effect (inner limit) and
the maximum greenhouse effect (outer limit).  Concerning the latter
it is assumed that a cloud-free CO$_2$ atmosphere
shall still be able to provide a surface temperature of 273~K.
By contrast, the inner limit of the
CHZ is defined by the onset of water loss.  In this case, a
wet stratosphere is assumed to exist where water is lost by
photodissociation and subsequent hydrogen escape to space.
Furthermore, the outer limit of the CHZ is defined by the first
CO$_2$ condensation attained by the onset of formation of CO$_2$
clouds at a temperature of 273~K; see, e.g., \cite{unde03}
for additional details and applications.

Owing to the shape of the photospheric spectra and the total
amount of the radiative energy fluxes, the limits of the habitable
zones are known to depend both on the stellar effective temperatures
$T_{\rm eff}$ and the luminosities $L_\ast$.  The results for the
CHZs and GHZs as well as the appropriate
Earth-equivalent positions for main-sequence stars between spectral
type F0 and G0 are given in Fig.~1.  For F0~V stars, the CHZ extends
between 2.27 and 2.92~AU and the GHZ extends between 1.99 and 3.67~AU.
Furthermore, for F8~V stars, the CHZ extends between 1.29 and 1.80~AU and
the GHZ extends between 1.14 and 2.21~AU.  The corresponding stellar data
are given in Table~1.  They are those also adopted for the photospheric
models computed with the PHOENIX code (see Sect. 3.1), which have
subsequently been used for our astrobiological studies.

Another aspect of habitability concerns the limits of the climatological
habitable zones for the various types of stars.  They have been
calculated based on the formalism by \cite{sels07}.  It provides
a suitable polynomial fit and, furthermore, implements the required
correction for the solar effective temperature in consideration of
that \cite{kast93} used for an unusually low value of 5700~K instead
of 5777~K as currently accepted.


\subsection{Impact of UV: The DNA Action Spectrum}

Next we comment on the relevance of the DNA action spectrum.  The most
fundamental radiometric technique to quantify radiative damage on
biomolecules and microorganisms is spectroradiometry.  Biological effectiveness
spectra can be derived from spectral data by multiplication with an action
spectrum $S_\lambda(\lambda)$ of a relevant photobiological reaction with the
action spectrum typically given in relative units normalized to unity for, e.g.,
$\lambda = 300$~nm.  The biological effectiveness for a distinct range of
the electromagnetic spectrum such as UV radiation is determined by
\begin{equation}
E_{\rm eff} \ = \
\int_{\lambda_1}^{\lambda_2}E_\lambda(\lambda)S_\lambda(\lambda)\alpha(\lambda)d\lambda \ \ ,
\end{equation}
where $E_\lambda(\lambda)$ denotes the stellar irradiance (ergs~cm$^{-2}$~s$^{-1}$~nm$^{-1}$),
$\lambda$ the wavelength (nm), and $\alpha(\lambda)$ the planetary atmospheric attenuation
function; see \cite{horn95}.  Here $\lambda_1$ and $\lambda_2$ are the limits of integration
which in our computations are set as 200 nm and 400 nm, respectively.
Although a significant amount of stellar radiation exists beyond 400 nm, this
portion of the spectrum is disregarded in the following owing to the minuscule
values for the action spectrum $S_\lambda(\lambda)$ in this regime.
Planetary atmospheric attenuation, sometimes also called extinction, results in a  
loss of intensity of the incident stellar radiation.  In Eq.~(1) $\alpha$ = 1 indicates
no loss and $\alpha$ = 0 indicates a complete loss (see Sect. 2.3); note that
$\alpha(\lambda)$ can be attained by various types of methods, which may also
consider detailed atmospheric photochemical models.

Based on previous work, the UV region of the electromagnetic spectrum has been
divided into three bands termed UV-A, UV-B, and UV-C.  The subdivisions are
somewhat arbitrary and differ slightly depending on the discipline
involved\footnote{Wavelengths above 380 nm are often considered already belonging
to the optical regime, which is however inconsequential for this study in consideration
of the very small values of the employed action spectrum between 360 and 400 nm.}.
Here we will use: UV-A, 400-320 nm; UV-B, 320-290 nm; and UV-C, 290-200 nm.
Following \cite{diff91}, the division between UV-B and UV-C is chosen as 290
nm since UV at shorter wavelengths is unlikely to be present in terrestrial
sunlight, except at high altitudes \citep{hend77}.  The choice of 320 nm as
the division between UV-B and UV-A is suggested by the level of photobiological
activity although subdivisions at 330 or 340 nm have previously also
been advocated \citep{peak86}.

In order to compute the irradiance toward targets in circumstellar environments,
typically positioned in stellar habitable zones, a further equation is needed,
which is
\begin{equation}
E_\lambda(\lambda) \ \propto \ F_\lambda (R_\ast/d)^2 \ \ ,
\end{equation}
where $F_\lambda$ is the stellar
radiative flux, $R_\ast$ is the stellar radius and $d$ is the distance between the
target and the star.  In the framework of this work, we will focus on planets
at different positions in stellar HZs.  For stellar UV radiation we will consider
photospheric radiation only (see Sect. 3) because the chromospheric UV radiation
from F-type stars is of minor importance \citep{lins80}.

Action spectra for DNA, also to be viewed as weighting functions, have previously
been utilized to quantify and assess damage due to UV radiation \citep{setl74}.
Besides DNA, action spectra have also been derived for other biomolecules,
for biostructures such as cellular components as well as for distinct species,
especially extremophiles \citep[e.g.,][and references therein]{horn95,rett02}.
\cite{horn95} provides information on the DNA action spectrum for the range from
285~nm (UV-C) to 400~nm (UV-A).  She found that between 400 and 300~nm, the
action spectrum increases by almost four orders of magnitude (see Fig.~2).
The reason for this behavior is the wavelength-dependency of the absorption and
ionization potential of UV radiation in this particular regime.  A further
significant increase in the DNA action spectrum occurs between 300 and 200~nm
(see below).  Fortunately, however, the Earth's ozone layer is very sufficient
to filter out this type of lethal radiation \citep[e.g.,][]{diff91,cock98}, which
in our models is mathematically dealt with by considering an appropriate
attenuation function $\alpha(\lambda)$ (see Eq.~1).

The behavior of the DNA action spectrum in the wavelengths regime between 200
and 290 nm has been given by \cite{cock99}.  He points out that a significant
reason for the susceptibility to UV-induced damage are $\pi$-electron systems,
notably because of their $\pi$ to $\pi^\ast$ energy transitions.  It is found
that this type of interaction also accounts for protein damage and damage to
enzymes of photosynthesis, which indicates that these types of biochemical
reactions are of general importance.  The increase in the relative biological
damage is about a factor of 35 relative to the damage at the reference
wavelengths of 300 nm (see Fig.~2).


\subsection{Planetary Atmospheric Attenuation}

A relevant ingredient to our study is the consideration of planetary atmospheric
attenuation $\alpha(\lambda)$, which typically results in a notable reduction
of the received stellar radiation.  Appropriate values for $\alpha(\lambda)$
can be obtained through the analysis of theoretical exoplanetary models
\citep[e.g.,][]{mead11,kalt12}, inspired by recent results from the {\it Kepler}
mission, or the usage of historic Earth-based data \citep[e.g.,][]{cock02}.
Within the scope of the present work that is mostly
focused on the impact of photospheric radiation from stars of different spectral
types, as well as on the role of active and inactive stellar chromospheres, we
assume that $\alpha(\lambda)$ is given by a parameterized attenuation function
ATT defined as
\begin{equation}
{\rm ATT}(\lambda) \ = \ {C \over 2}~\Bigl[ 1 + \tanh (A (\lambda-B)) \Bigr] \ \ ;
\end{equation}
see Fig.~3 for a depiction of different examples $(A, B, C)$.  Here $A$ denotes
the start-of-slope parameter, $B$ (in nm) the center parameter, and $C$ the maximum
(limited to unity) of the distribution.


For example, \cite{cock02} provided information about the ultraviolet irradiance
reaching the surface of Archean Earth for various assumptions about Earth's
atmospheric composition; the latter allow us to constrain the wavelength-dependent
attenuation coefficients for Earth 3.5 Gyr ago.  Furthermore, there is a large array
of recent studies about exosolar planetary atmospheres, including those for rocky planets.
They encompass models regarding the detailed treatment of atmospheric photochemistry,
including the build-up and destruction of ozone, as discussed by, e.g., \cite{segu03},
\cite{scal07} and \cite{gren07}.  These models provide information on different
exoplanetary structures due to outside forcings, including variable stellar radiation,
which in principle allow the derivation of detailed planetary atmospheric attenuation
functions.


\section{Stellar Evolution, Photospheric Models and Irradiance}

\subsection{Basic Properties}

In accordance to previous studies of our group, as given by \cite{bloh09}
and \cite{cunt12}, which mostly focused on super-Earth planets, we rely
again on selected stellar evolution models that have been computed with
the well-tested Eggleton code; see, e.g., \cite{schr08} for a description
of an adequate solar evolutionary model.  The Eggleton code allows us to
take into account the changing properties of the host star during its
evolution on the main-sequence and, subsequently, as a red giant.  We use
an advanced version of the Eggleton code, including updated opacities
and an improved equation of state as described by \cite{pols95,pols98}. 
Besides other desirable characteristics, the adopted evolution code utilizes
a self-adapting mesh and also permits treating ``overshooting" --- a concept 
of extra mixing, which has been thoroughly tested considering observational 
constraints.  In particular, the two convection parameters, the 
``mixing length'' and the ``overshoot length'', have been calibrated by 
matching accurately the physical properties of various types of stars, 
including giants and supergiants of well-known masses, found in well-studied, 
eclipsing binary systems \citep{schr97}.

For the abundance of heavy elements, which decisively affect the opacities,
we use the near-solar value of $Z=0.02$.  This choice is an appropriate
representation of present-day samples of stars in the thin galactic disk, 
noting that they exhibit a relatively narrow distribution ($Z=0.01$
to 0.03) about this value.
The adopted evolution code also considers a detailed and well-tested
description of the stellar mass loss, which becomes important for the final 
stages of giant star evolution; see \cite{schr05} and \cite{schr07}.  
Regarding our models of planetary host stars, the principal input parameters
are  the total luminosity and effective temperature, which are found to change
with time. Obviously, the resulting total lifetime of the star 
is a quantity of high significance as well.

While the effective temperature and luminosity of the host star already
allow a good representation of circumstellar habitability, as done through
the stellar climatological habitable zone (see Sect. 2.1), the irradiation
especially in the UV regime is of pivotal importance as well for
arriving at a realistic evaluation of habitability.  The main
constraint arises from having sustainable conditions for biological
organisms and biochemical processes, which provide the basis of life.
Here, damaging ultraviolet radiation must be of particular concern.  Its
share regarding the total stellar luminosity critically depends on the
stellar effective temperature and increases significantly from late (F9)
to early (F0) F-type stars.

We employ the necessary and accurate account of stellar radiation, including 
its spectral energy distribution, by utilizing a number of photospheric
models computed by the PHOENIX code following \cite{haus92}; see Fig.~4.  
The adopted range
of models for the F-type stars are in response to effective temperatures
of 7200~K for spectral type F0, 7000~K for F1, 6890~K for F2, 6700~K 
for F3, 6440~K for F5, 6200~K for F8, and 6050~K for G0.  The
PHOENIX code iterates the principal physics and structure of a stellar
atmosphere until a final model is obtained, which is in radiative and
hydrostatic equilibrium; see \cite{haus03}.  As part of this procedure, energy
transport by convection is also considered.  The PHOENIX code solves the
equation of state, including a very large number of atoms, ions, and
molecules.  With respect to radiation, the one-dimensional spherically symmetrical
radiative transfer equation for expanding atmospheres is solved, including a
treatment of special relativity; see \cite{haus99b}. Opacities are sampled
dynamically over about 80 million lines from atomic transitions and
billions of molecular  lines, in addition to background (i.e., continuous)
opacities, to provide an emerging spectral flux as reliable and realistic 
as possible.

As part of our study, all spectra emergent from a stellar model atmosphere
have first been calculated with a high resolution of $\Delta\lambda=0.01\,$\AA,
for a highly complete inclusion of the lines.  The spectra were then binned
down to a much lower resolution, which is more practical for our subsequent
astrobiological analyses.


\subsection{Evolution of F-type Stars}

Main-sequence stars, including the Sun, represent the slowest phase
of stellar evolution, since at that stage the largest energy reservoir
of the star is consumed: its central hydrogen is converted to helium;
see Fig.~5 for examples of F-star evolutionary tracks.  All later
phases, where the star becomes a red giant twice, before and after
the ignition of the central helium burning, are much faster and
present much larger changes in stellar luminosity; see \cite{cunt12}
for a detailed study of the implications for habitable super-Earths.
Hence, as previously found, main-sequence stars are, in general,
most promising in the context of astrobiology.  Though there are limits,
since an increasing  mass accelerates stellar evolution considerably.
Nonetheless, F-type stars corresponding to the mass-range of about
1.1 to 1.6 $M_\odot$, provide stable lifetimes of 2 to 4 billion
years, expected to be sufficient for the origin and evolution of life.

However, regarding the evolution of effective temperatures and, in
consequence, of the spectral distribution of the emergent radiation,
F-type stars differ somewhat from the Sun, especially the most massive
ones:  Their effective temperatures rise to much higher values while
still staying on the main-sequence, and there is a quick fall afterwards,
and a rise again toward the end of this phase.  The reason for this
evolutionary behavior resides in the cores of F-stars:  while the solar
core is not convective but relies purely on radiative energy transport,
F-stars employ the more efficient process of convection for transporting
their high amounts of produced energy to the outer layers.  Inside the
most massive F-stars, where core convection has gained full power,
rising bubbles even overshoot the boundary set by the Schwarzschild
criterion.  Hence, as a by-product, the convective cores of F-stars benefit 
from an enhanced chemical mixing by gaining access to a larger hydrogen 
reservoir around them, within reach of the ``overshoot length''.

Therefore, F-type stars still spend a relatively long time on the
main-sequence,  making good in part for their faster evolution in general.
This phenomenon is most notable between 1.4 and 1.5 $M_\odot$, where
overshooting sets in.  However, the evolutionary behavior of F-stars after
the end of central hydrogen burning also differs a bit from that of the Sun,
which  does neither experience overshooting nor any core convection.
These particulars of stellar evolution determine both the stellar effective
temperatures and luminosities (see Table~1); they also drive the changes
of the spectral characteristics of the irradiation and, finally, the inner
and outer limit of the respective habitable zones.

Compared to G-type and K-type stars, F-type stars move relatively fast
through the final stages of evolution beyond the main-sequence.  In those
stages, very dramatic changes in luminosity (increasing by over a factor
10$^3$), stellar radius (increasing by over  a factor 100) and effective
temperature (decreasing by over a factor 2) occur.  The corresponding
time scales are limited to a few hundred million years compared to the
Sun, which will take about 2 billion years for completing these steps.
For these reasons, we will restrict the study of circumstellar habitability
of F-stars to their phases on and near the main-sequence.


\subsection{Evolution of F-star Climatological Habitable Zones}

Next we focus on the evolution of F-star climatological habitable zones,
and the subsequent identification of the inner and outer limits of the
CHZs and GHZs.  As pointed out in Sect. 3.2, F-type stars, alike the Sun,
encounter a slow but consistent growth of luminosity during their
main-sequence phase as well as characteristic changes of the stellar
effective temperatures.  This is relevant for all types of F-stars,
i.e., for the entire range of masses between about 1.2 and 1.5 $M_\odot$
as addressed in the following.  These masses are identified as such at
the zero-age main-sequence (ZAMS), and they experience little change
prior to the departure of the star toward the red giant domain;
the latter is outside the scope of this study.

An example is given as Figure 6.  It shows the evolution of the
climatological habitable zone for a star with an initial mass of 1.3~$M_\odot$.
It also depicts the development of the Earth-equivalent position\footnote{
A more technical definition of an Earth-equivalent position involves an
interpolation between the inner and outer limits of the CHZ. In this case,
it depends both on the stellar luminosity and effective temperature;
see, e.g., \cite{sels07} and \cite{cunt14} for details.} given
as $\sqrt{L_\ast}$ (with $L_\ast$ as stellar luminosity in solar units),
which is progressively moving outward.  Note that by the
end of central hydrogen burning, the total stellar radiation output
has almost doubled, entailing a considerable increase of both the
inner and outer limits of the climatological habitable zone,
identified either as CHZ or GHZ.  Furthermore, the stellar spectral
types change due to alterations of their effective temperatures
during the course of stellar evolution, including main-sequence
evolution.  In the latter case, stars of initial masses between
1.2 and 1.5 $M_\odot$ stay within the F-type range (see Table 2).

Although it is appropriate to explore UV-based habitability at any
distance from a planetary host star, we rather select the inner and 
outer limits of the CHZ (labelled as iC and oC, respectively), and
the inner and outer limits of the GHZ (labelled as iG and oG,
respectively) as venues of our study.  Results are given in Table 3
and Figure 7.  We especially consider the extreme positions of the
iG, iC, oC, and oG attained as consequence of the stellar
main-sequence evolution, i.e., as identified at the ZAMS stage and
when the stars depart from the main-sequence.  Our study indicates
once again that when stars age the climatological habitable zones
become broader and migrate outwardly.

A special aspect of significance to astrobiology concerns the
continuous domains of both the CHZs and GHZs, also referred
to as continuously habitable zones (see Fig. 7), where the criteria
for the CHZs and GHZs (see Sect. 2.1) are fulfilled for a distinct
period of time (here: the stellar main-sequence stages).  The
continuously habitable zones have evidently much smaller widths than
the habitable zones defined by the extremes for the same amounts of
time.  The continuous domains of habitability for our set of stars
are particularly small for the CHZs.  For a 1.2 $M_\odot$ this domain
extents from 1.366 to 1.688 AU, whereas for a 1.5 $M_\odot$ it extents
only from 2.535 to 2.566 AU; this aspect should be taken into account
for any comprehensive habitability assessment.


\section{Results and Discussion}

\subsection{Stellar Case Studies}

The aim of our study is the investigation of biological damage inflicted upon DNA
due to stellar UV radiation for objects around main-sequence F-type stars.  Our results
will be given in terms of $E_{\rm eff}$, a parameter defined as the ratio of damage for
a given distance from the star for an object with or without atmospheric attenuation
relative to the damage for an object at 1~AU from a solar-like star  without atmospheric
attenuation present.  We will focus on the following aspects: (1) influence of spectral
types of the host stars, especially for F0~V, F2~V, F5~V, and F8~V, (2) relevance of
planetary positions in the stellar habitable zones, including the general and
conservative inner limits, Earth-equivalent positions, and the general and
conservative outer limits, (3) effects of planetary atmospheric attenuation
approximated by analytic functions, and (4) the relative importance of UV-A, UV-B,
and UV-C regarding the DNA damage.

The DNA damage in the habitable zones around F0~V, F2~V, F5~V, and F8~V stars
in cases without planetary atmospheric attenuation
is found to be almost always more significant than for the reference object at
Earth distance from a solar-like star (see Figure 8).  The sole exception is for
objects at the outer limit of the GHZ around an F8~V star, which
in the solar system would correspond to an orbit beyond Mars; here $E_{\rm eff}$ is
given as 0.95.  Planetary atmospheric attenuation effectively reduces the damage of DNA,
as expected (see Figure 9).  In this work, as a standard example adopted for tutorial
reasons, the planetary attenuation parameters are chosen as $A = 0.05$, $B = 300$,
and $C = 0.5$, except otherwise noted.  In this case of attenuation, the damage on
DNA, for positions at the inner limit of the GHZs, is drastically reduced compared
to the reference object without any attenuation.  For other positions in the stellar
habitable zones, the reductions are relatively high as well (see Figure 9).

It is particularly intriguing to compare the respective values of $E_{\rm eff}$ for
F0~V, F2~V, F5~V, and F8~V stars for different positions in the stellar habitable
zones as they indicate the role of the significantly higher UV fluxes of F-type stars
compared to the photospheric radiation of a solar-like star.  For the inner limits
of the GHZs (i.e., the closest positions to the stars considered),
$E_{\rm eff}$ for F0~V, F2~V, F5~V, and F8~V stars are identified as 9.8, 7.3, 4.2,
and 3.6 without atmospheric attenuation; they decrease to 0.23, 0.19, 0.14, and 0.14,
respectively, when our default choice of atmospheric attenuation is applied.
At Earth-equivalent positions (which depend on the stellar spectral type much like 
the above-mentioned limits of habitability), $E_{\rm eff}$ for each star are given 
as 7.1, 5.2, 3.0, and 2.5, respectively, without atmospheric attenuation, and 
as 0.16, 0.14, 0.10, 0.10 with attenuation.  At the outer limits of GHZs, $E_{\rm eff}$
for each star is given as 2.9, 2.1, 1.1, and 0.95 without atmospheric attenuation,
and as 0.067, 0.055, 0.038, and 0.037 with attenuation.  Intermediate values are
obtained for the inner and outer limits of the CHZs.

In all cases DNA damage due to UV-A, compared to UV-B and UV-C, is minuscule at best.
$E_{\rm eff}$ due to UV-A at the inner limit of the GHZ for an F0~V star is
$2.46 \times 10^{-5}$ without atmospheric attenuation, which is the maximum value
of all cases considered.  This value is further reduced to $1.13 \times 10^{-5}$
in case of default atmospheric attenuation.  Of the three regimes of UV, the one most
affected by our above choice of planetary attenuation is UV-C.  For F0~V, F2~V,
F5~V, and F8~V, the DNA damage due to UV-C is 96\%, 95\%, 93\%, and 91\%, respectively,
of the total for the UV regimes combined.  If atmospheric attenuation is included, the
DNA damage due to UV-C is reduced to 1.7\%, 1.9\%, 2.3\%, and 2.6\% of the damage due
to UV-C without planetary attenuation.  It comprises 70\%, 68\%, 65\%, and 61\% of the
total for each case of attenuation.  The relative damage attributable to UV-C decreases
with stellar types from F0~V to F8~V.  The damage due to UV-A and UV-B also decreases
with stellar types from F0~V to F5~V, but it slightly increases between F5~V and F8~V
because of the shape of the stellar photospheric spectrum (see Figure 5).

\subsection{Impact of Planetary Atmospheric Attenuation}

Since atmospheric attenuation is expected to impact preferably the UV regime,
which in turn is most relevant to the DNA damage, we also explored the effects
of parameter choices for the attenuation function ATT (see Eq.~3).
For tutorial reasons, we focused on the F2~V star, and varied
one of the parameters, i.e., $A$, $B$, or $C$, at a time.  Figure 10 depicts
two cases out of several combinations of fixed parameters that were investigated.
In the top panels of Figure 10, the parameter $A$ is varied from 0.01 to 1.0, whereas
the other parameters remain fixed at $B$ = 300 and $C$ = 0.5.  In the bottom panels,
the parameter $B$ is varied from 250 to 350 and the two other parameters remain
fixed at $A$ = 0.05 and $C$ = 0.5.  The two panels on the left show the relative
impact of UV-A, UV-B, and UV-C on DNA at an Earth-equivalent position.  Note that
the impact of UV-A is, however, completely unrecognizable in all panels.  The two
panels to the right show the DNA damage at five selected positions in the habitable
zone of the F2~V star, which are: the inner limits of the GHZs and CHZs,
the Earth-equivalent position, and the outer limits of the CHZs and GHZs.
It is found that damage inflicted on
DNA is considerably larger for objects closer to the star.  Thus, in Fig.~10,
the top two lines represent the inner limits of the CHZ and GHZ, and the bottom
two lines represent the outer limits of the CHZ and GHZ.

The parameter $A$ depicts the rate of change in ATT with increasing wavelength
(see Fig.~3).  Thus, a smaller value of $A$ corresponds to a gentler slope of
ATT, but also to a considerably higher value of ATT at relatively short wavelength
(i.e., near 200~nm), which implies a higher effect of attenuation where damage
inflicted on DNA is most severe (see Fig.~2).  The parameter $B$ describes the
center wavelength of the attenuation function; it also changes the response of
each regime to varied values of $A$.  Thus, the appearance of $E_{\rm eff}$ as
function of the attenuation parameter $A$ heavily depends on the fixed value of
parameter $B$.  For example, let us consider the case $B = 300$.  Here, the DNA
damage due to UV-A increases, while the damages due to UV-B and UV-C decrease
as the parameter $A$ increases (see Fig. 10).  Here $E_{\rm eff}$ decreases from
0.73 to 0.0074 for an Earth-equivalent position with increasing $A$.  The largest
changes occur between $A = 0.01$ and 0.08, where $E_{\rm eff}$ due to UV-C declines
drastically.  In these cases, especially for our standard choice of attenuation 
(see Sect. 4.1), most of the DNA damage is attributable to UV-C.  But for
$A = 0.08$, $E_{\rm eff}$ due to UV-B and due to UV-C are about the same, and
for larger values of $A$, $E_{\rm eff}$ due to UV-B becomes greater than 
$E_{\rm eff}$ due to UV-C.

We also varied the parameter $B$ while keeping the values for $A$ and $C$ fixed.
$E_{\rm eff}$ is identified as a decreasing function for variable values of $B$.
Since parameter $B$ determines the center of the attenuation function, the effect
of UV diminishes from the lowest wavelengths as $B$ is increased.  In other words,
UV-C is affected the most by varying $B$.  If $A = 0.05$ and $C = 0.5$ are used,
$E_{\rm eff}$ decreases from 1.32 at $B = 250$ to 0.14 at $B = 300$ and to 0.0012
at $B = 350$ for objects at Earth-equivalent positions.  This behavior is shown
in Figure 10.  If $A = 0.1$ and $C = 0.5$ are used, the appearance of each
functional dependence is similar to the case of $A = 0.05$; however, $E_{\rm eff}$
of UV-B exceeds that of UV-C at $B = 293$.  The functional dependence for UV-C
is steeper for $A = 0.1$ than for $A = 0.05$.  The total amount of $E_{\rm eff}$
decreases from 1.36 at $B = 250$ to 0.048 at $B = 300$ and to $3.9 \times 10^{-6}$
at $B = 350$ for objects at the Earth-equivalent position.

DNA damage is inversely proportional to the square of distance between
a planet and the host star, and thus, the ratio of DNA damage (with or
without the consideration of atmospheric attenuation) at one position to
another does not change with the selected combination of parameters.
Because of that, if $E_{\rm eff}$, the ratio of DNA damage at a position
to the reference value, at different positions are compared, the differences
are smaller for more effective combinations of parameters.  For example,
the DNA damage at the inner limits of the GHZs and the CHZs, and at the
outer limits of the CHZs and GHZs are 139\%, 108\%, 62\%, and 40\% of
DNA damage as measured by $E_{\rm eff}$, respectively, as found for the 
F2~V star.  If attenuation is considered, and if the attenuation parameters
are chosen as $(A, B, C) = (0.01, 300, 0.5)$, $E_{\rm eff}$ is found as
1.01, 0.78, 0.73, 0.45, and 0.29 at the inner limit of the GHZ and CHZ, 
at the Earth-equivalent position, and at the outer limit of the CHZ and
GHZ, respectively.  By contrast, if the same parameters are set as
$(A, B, C) = (0.5, 300, 0.5)$, our standard choice), $E_{\rm eff}$ is
identified as 0.0119, 0.0092, 0.0086, 0.0053 and 0.0034, respectively.


\subsection{Habitability during the Course of F-Star Evolution}

We also studied the influence of stellar main-sequence evolution, resulting
in pronounced changes in the stellar effective temperatures and luminosities
regarding the damage inflicted on DNA.  Again, we consider cases without and  
with planetary atmospheric attenuation (see Sect. 4.1).
Specifically, we explore the change in
$E_{\rm eff}$ at specific positions in the habitable zones for stars with
masses of 1.2, 1.3, 1.4, and 1.5 $M_\odot$.  The selected positions are the
general outer limit at ZAMS (bottom dotted line in Figures 11 and 12), the
conservative outer limit at ZAMS (bottom dashed line), the general inner limit
at the end of main-sequence (top dotted line), the conservative inner limit
at the end of main-sequence (top dashed line), the Earth-equivalent positions
at ZAMS (i.e., minimum distance; top solid line) and the end of main-sequence
(i.e., maximum distance value; bottom solid line); see Sect. 3.3.  The distances
for various positions are given in Table 4 and Figure 7.

Furthermore, we also consider average (i.e., time-independent) Earth-equivalent
positions, which are derived by interpolating the conservative outer limit at
the ZAMS and the conservative inner limit at the end of main-sequence evolution
noting that these limits correspond to the continuous domains of the CHZs.
Generally, the Earth-equivalent positions are computed as weighted averages
between the inner and outer limits of the CHZs, which can be approximated as
$0.95 \sqrt{L_\ast}$ and $1.37 \sqrt{L_\ast}$ (with $L_\ast$ in solar units),
respectively, with the Earth-equivalent positions given as $\sqrt{L_\ast}$
\citep{kast93}.  For stars with masses of 1.2, 1.3, 1.4, and 1.5 $M_\odot$,
the average Earth-equivalent positions are obtained as 1.56, 1.82, 2.10, and
2.54~AU, respectively, which in the Solar System would correspond to an approximate
distance nearly between Mars and Ceres.  Moreover, we also consider evolving
Earth-equivalent positions, noting that the inner and outer limits of CHZs
change on evolutionary time scales (see Fig.~7); see discussion below.

In principle, the damage inflicted on DNA at any position within the stellar
habitable zones at the ZAMS increases as a function of the stellar mass
(see Figures 11 and 12).  Regarding the inner and outer limits of both the
CHZs and GHZs, the damage on DNA first increases with evolutionary time,
but then starts decreasing with time, while the stellar luminosity keeps
increasing and both the CHZs and GHZs continue to migrate outward.  If
no planetary attenuation is considered, the ZAMS $E_{\rm eff}$ values at
average Earth-equivalent positions for stars of 1.2, 1.3, 1.4, and
1.5 $M_\odot$ are found as 0.91, 2.46, 3.41, and 5.02, respectively.
The maximum values for the DNA damage, expressed as $E_{\rm eff}$, at
average Earth-equivalent positions are given as 1.96, 2.63, 3.48, and 5.07,
respectively.  These values are obtained at stellar evolutionary times
of 1.95, 0.93, 0.27, and 0.14 Gyr, respectively; it means that the
maximum amounts of damage inflicted on DNA are attained much earlier for
stars of higher mass (and, by implication, of earlier spectral type and
higher initial effective temperature).  At the end of main-sequence
evolution, the $E_{\rm eff}$ values at average Earth-equivalent positions
are then reduced to 0.96, 1.04, 1.07, and 0.32, respectively; they are
comparable to or smaller than those for a reference object orbiting
a solar-like star.

If default planetary atmospheric attenuation is assumed, corresponding
to $(A, B, C) = (0.05, 300, 0.5)$, the damage is drastically reduced,
i.e., by up to 96\%, 97\%, 97\%, and 98\% for stars of 1.2, 1.3,
1.4, and 1.5 $M_\odot$, respectively.  The precise amount of reduction
is a function of the stellar evolutionary status.  The amount of
reduction moreover depends on the type of star, which determines the
shape of the emergent spectrum, therefore entailing different amounts
of damage.  As previously discussed, the damage is dominated by the
UV-C regime (see Eq. 1 and Fig. 2).
If standard attenuation is considered, the ZAMS values for $E_{\rm eff}$
at average Earth-equivalent positions for stars of 1.2, 1.3, 1.4, and 1.5
$M_\odot$ are reduced to 0.038, 0.082, 0.098, and 0.121, respectively.
The maximum $E_{\rm eff}$ values at those positions are given as
0.075, 0.087, 0.099, and 0.122, whereas at the end of the main-sequence
stages, they are further reduced to 0.040, 0.041, 0.041, and 0.014,
respectively.

For specific age intervals, $E_{\rm eff}$ changes in different ways
in the vicinities of stars of different masses.  The following data
refer to objects at average Earth-equivalent positions, but they can
also be converted for other star--planet distances in a highly
straightforward manner.  We are particularly interested in the change
of $E_{\rm eff}$ between 0.5 Gyr and 2.5 Gyr, i.e., during the early
stages of the systems, contemporaneous with the time when life
originated on Earth \citep[e.g.,][and references therein]{brac98},
For a star with masses of 1.2 $M_\odot$, $E_{\rm eff}$ increases from
1.75 to 1.96 until 1.95 Gyr, and thereafter decreases to 1.92 at the
average Earth-equivalent position if planetary atmospheric attenuation
is absent.  For a star with masses of 1.3 $M_\odot$, $E_{\rm eff}$
increases from 2.61 to 2.63 until 0.93 Gyr, and then decreases to
1.86.  For stars with masses of 1.4 $M_\odot$ and 1.5 $M_\odot$,
$E_{\rm eff}$ decreases from 3.46 to 1.10 and decreases from 4.88 to
0.52, respectively.  Thus, a planet in the habitable zone of a
1.2 $M_\odot$ mass star will experience the least change in the
amount of DNA damage, whereas that change will be greatest for a 
1.5 $M_\odot$ mass star.

Adequate models should also take into account effects due to planetary
atmospheric attenuation.  In this case, even the climatological
habitable zone of a 1.5 $M_\odot$ mass star may be able to offer
a relatively well UV protected environment throughout the early
2 Gyr period.  The same statement is expected to apply for F-type
stars of lower mass as well.  If our default model of atmospheric
attenuation is adopted, $E_{\rm eff}$ for a 1.2 $M_\odot$ mass star
increases from 0.069 to 0.075 until 1.95 Gyr, and then decreases to
0.074; note that these data refer to average Earth-equivalent positions. 
For a 1.3 $M_\odot$ mass star, $E_{\rm eff}$ for the same setting
increases from 0.086 to 0.087 until 0.93 Gyr, and then decreases
to 0.067.  In comparison, for stars with masses of 1.4 and
1.5 $M_\odot$, $E_{\rm eff}$ decreases from the maximum value
of 0.099 to 0.042 and from 0.118 to 0.021, respectively.  For
other planetary positions, the respective values for $E_{\rm eff}$
can be obtained through appropriate scaling.

Moreover, we also computed the DNA damage at evolving (i.e.,
time-dependent) Earth-equivalent positions, depicted as dashed
lines in Fig.~7.  For all cases from 1.2 to 1.5 $M_\odot$, the
Earth-equivalent positions at ZAMS are located very close to the stars.
In fact, they are situated interior to the continuous domains of both
the CHZs and GHZs.  As the stars age, the Earth-equivalent positions
migrate outward, crossing or even passing the continuous domains of
the CHZs, especially for relatively massive F-type stars.  Since the
damage inflicted on DNA is proportional to the square of distance from
the star, the attained results reflect the behavior of the
planetary positions.  The maximum DNA damage at evolving Earth-equivalent
positions without atmospheric attenuation for stars of 1.2, 1.3, 1.4,
and 1.5 $M_\odot$ are given as 2.55, 3.56, 4.74, and 7.74, respectively,
occurring at very early stages of the stellar lifetimes, whereas
the minimum DNA damage are given as 0.92, 0.99, 1.01, and 0.29, respectively,
obtained at the end of main-sequence evolution (see Fig.~11).  Damage
on DNA is reduced by planetary atmospheric attenuation, as expected.
For default attenuation, the maximum damage is found as 0.099,
0.119, 0.136, and 0.186, respectively, whereas the minimum damage is
found as 0.038, 0.040, 0.039, and 0.012, respectively (see Fig.~12).


\section{Summary and Conclusions}

In this study, we investigated the general astrobiological significance
of F-type main-sequence stars.  DNA has been taken as a proxy for
carbon-based macromolecules following the paradigm that extraterrestrial
biology may most likely be associated with hydrocarbon-based biochemistry.
Consequently, the DNA action spectrum was utilized to describe the impact
of the stellar UV radiation.  We considered an array of important aspects,
including (1) the role of stellar main-sequence evolution, (2) the situation
for planets at different positions within the stellar habitable zones, and
(3) the general influence of planetary atmospheric attenuation, which has
been described based on a parameterized attenuation function.  The damage
on DNA was described by the output parameter $E_{\rm eff}$, defined as the
ratio of damage for a given distance from the star for a general object
relative to the damage for an object at 1~AU from a solar-like star
with attenuation absent.

We found the following results:

\smallskip\noindent
(1) For average Earth-equivalent planetary positions, located inside the
continuous domains of the CHZs, $E_{\rm eff}$ for stars of spectral type
F0~V, F2~V, F5~V, and F8~V are obtained as 7.1, 5.2, 3.0, and 2.5, respectively.
Earth-equivalent planetary positions depend on the stellar spectral type,
and are at greater distance for stars of higher effective temperature or, by
implication, larger mass.  These results are consistent with the earlier work by
\cite{cock99}.

\smallskip\noindent
(2) For the inner and outer limits of the CHZs as well as GHZs,
the results for $E_{\rm eff}$ can be obtained by scaling.  Specifically,
the damage inflicted on DNA is considerably increased at the inner limits of the
CHZs and GHZs and considerably decreased at the outer limits of the
CHZs and GHZs relative to average Earth-equivalent positions. For F2~V stars,
the scaling factors for the inner limits are given as 139\% and 108\%
and for the outer limits as 62\% and 40\% relative to that at average
Earth-equivalent positions.

\smallskip\noindent
(3) Owing to the form of the DNA action spectrum (see Fig.~2), in the absence of
significant planetary atmospheric attenuation, most of the damage on DNA is
because of UV-C.  Damage due to UV-B is significantly lower, and damage due to
UV-A is virtually nonexistent.  Regarding the latter, the largest value 
in the context of this study was attained at the inner limit of the GHZ
for an F0~V star in the absence of planetary atmospheric attenuation,
which is $2.46 \times 10^{-5}$.

\smallskip\noindent
(4) Planetary atmospheric attenuation, especially that associated with ozone
layers \citep[e.g.,][]{segu03}, is able to reduce damage inflicted on DNA
drastically.  In consideration of realistic atmospheric attenuation functions,
this aspect entails a drastic reduction of damage associated with UV-C.  On
a relative scale, it thus tends to increase the importance of UV-B.

\smallskip\noindent
(5) It is particularly intriguing to assess the behavior of $E_{\rm eff}$
during stellar main-sequence evolution, which has been evaluated in detail
for stars of masses of 1.2, 1.3, 1.4, and 1.5 $M_\odot$.  If no planetary
attenuation is taken into account, the ZAMS $E_{\rm eff}$ values at
average Earth-equivalent positions are identified as 0.91, 2.46, 3.41, and
5.02, respectively.  The $E_{\rm eff}$ values at other positions in the
stellar habitable zones can be obtained by scaling noting that the incident
stellar radiation is diluted following the inverse square law. 

\smallskip\noindent
(6) Taking average Earth-equivalent positions as reference, the $E_{\rm eff}$
values are found to change with time in response to changes in the stellar
parameters owing to the stellar main-sequence evolution.  $E_{\rm eff}$ first
increases with time and reach maximum values of 1.96, 2.63, 3.48, and 5.07
for stars with masses of 1.2, 1.3, 1.4, and 1.5 $M_\odot$, respectively; they
are attained at evolutionary times of 1.95, 0.93, 0.27, and 0.14 Gyr,
respectively.  Thereafter, the $E_{\rm eff}$ values decline.  They reach
0.96, 1.04, 1.07, and 0.32 for the selected set of stars.  These values are
comparable to or smaller than that at an Earth-equivalent position for a
solar-like star.

\medskip

Our study is a further contribution toward the exploration of the
exobiological suitability of stars hotter and, by implication, more massive than
the Sun.  Although these stars are relatively rare compared to G-type solar-type
stars, they possess significantly augmented habitable zones.  On the other hand,
their emergent photospheric UV fluxes are much larger; fortunately, however,
they can be diminished through planetary atmospheric attenuation.  Thus, at least
in the outer portions of F-star habitable zones, UV radiation should not be viewed
as an insurmountable hindrance to the existence and evolution of life.  Future
studies for F-type stars should encompass (1) detailed chemical models
of planetary atmospheres, aimed at constraining the attenuation parameters,
(2) examples of specific star--planet systems with information attained from
observational constraints, and (3) cases of F-type stars that are members of binary
(or higher order) systems.  Studies of the circumstellar habitability for those
systems also encompassing analyses of planetary orbital stability have been
given by, e.g., \cite{cunt14} and others.

\bigskip
\bigskip

\noindent
{\bf Acknowledgements.}
This work has been supported by the Department of Physics, University of
Texas at Arlington (S. S., M. C.), by a UGTO/PROMEP-funded project (D. J.), 
and by a CONACyT master student stipend (C. M. G. O.).  Additionally,
K. P. S. is grateful for CONACyT support of his sabbatical year projects
under application No. 207662.

\clearpage


\clearpage


\begin{figure*}
\centering
\begin{tabular}{c}
\epsfig{file=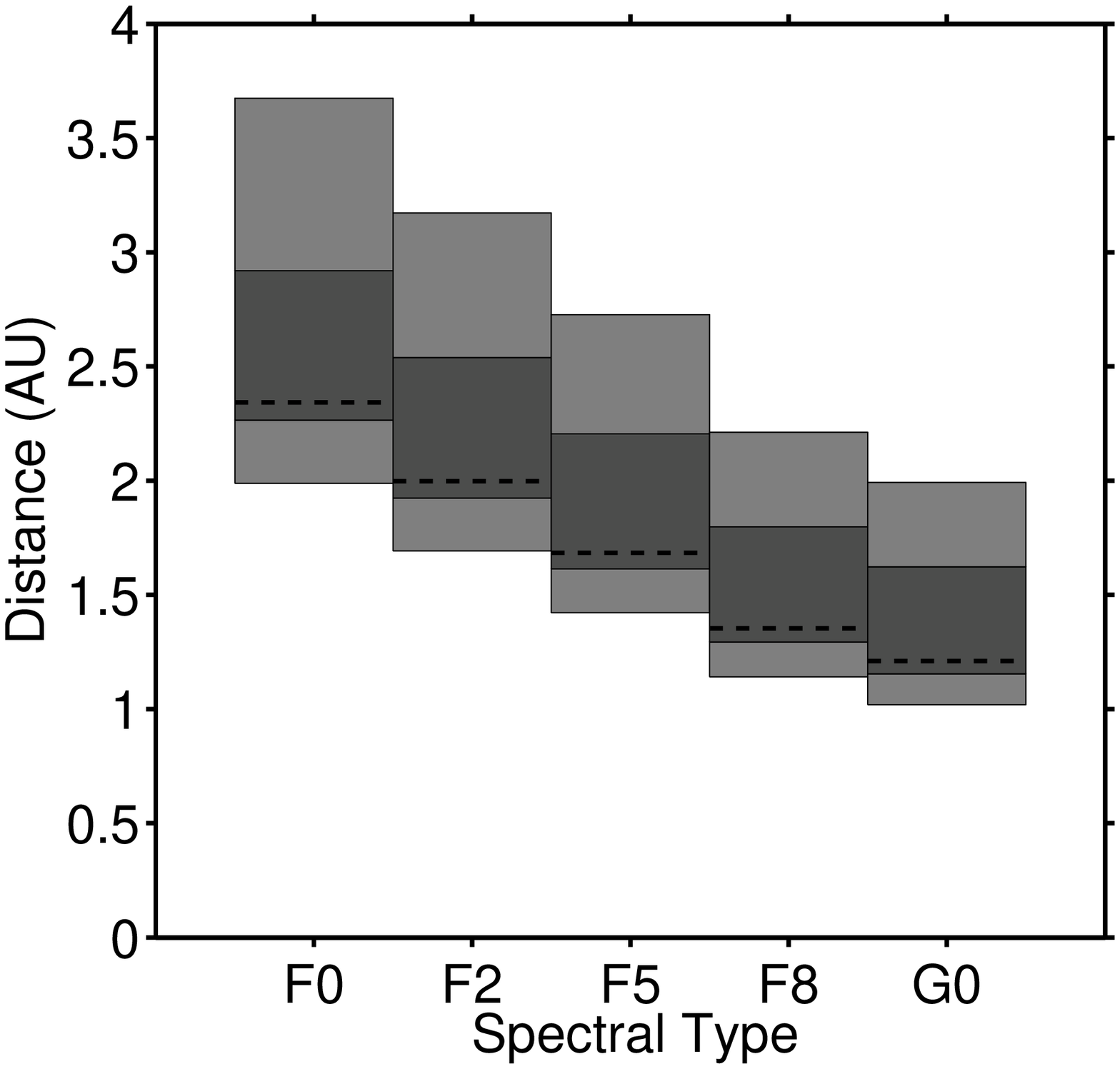,width=0.75\linewidth}
\end{tabular}
\caption{
Depiction of the habitable zones for F-type stars (see Table~1 for
information on the stellar parameters).  Dark gray colors indicate
the CHZs, whereas light gray colors indicate the GHZs.  Earth-equivalent
positions within the habitable zones are depicted by dashed lines.
}
\end{figure*}

\clearpage


\begin{figure*}
\vskip 5 cm
\centering
\begin{tabular}{c}
\epsfig{file=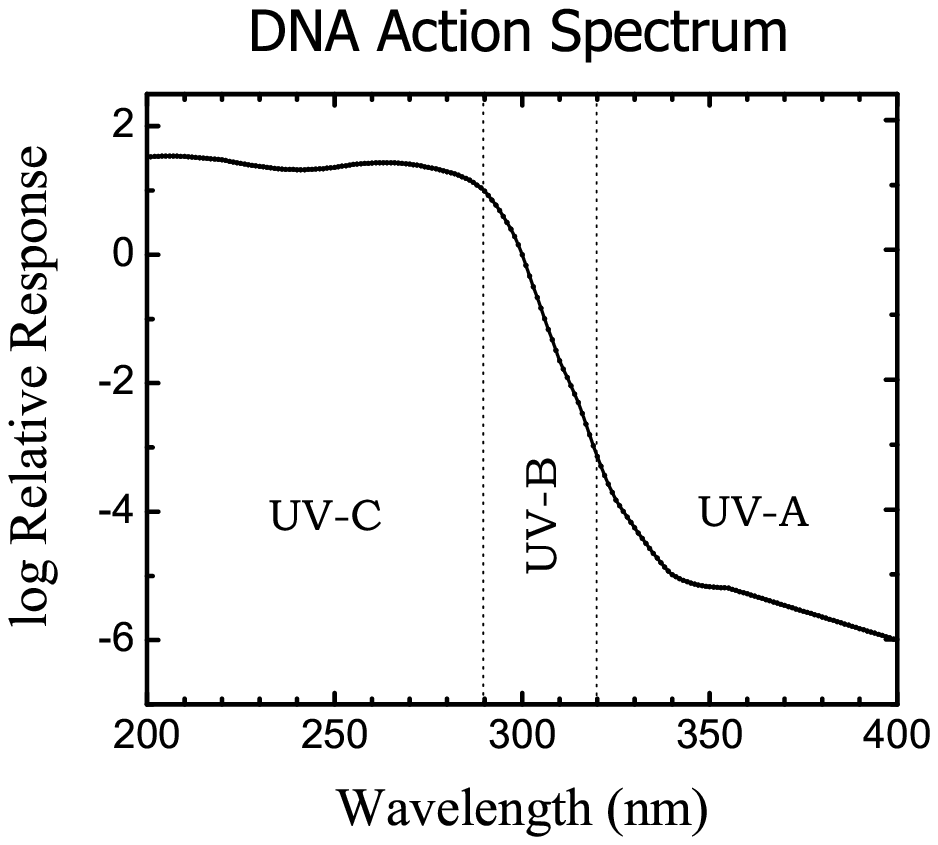,width=0.75\linewidth}
\end{tabular}
\caption{
DNA action spectrum following \cite{horn95} and \cite{cock99}.
}
\end{figure*}

\clearpage


\begin{figure*}
\centering
\begin{tabular}{c}
\epsfig{file=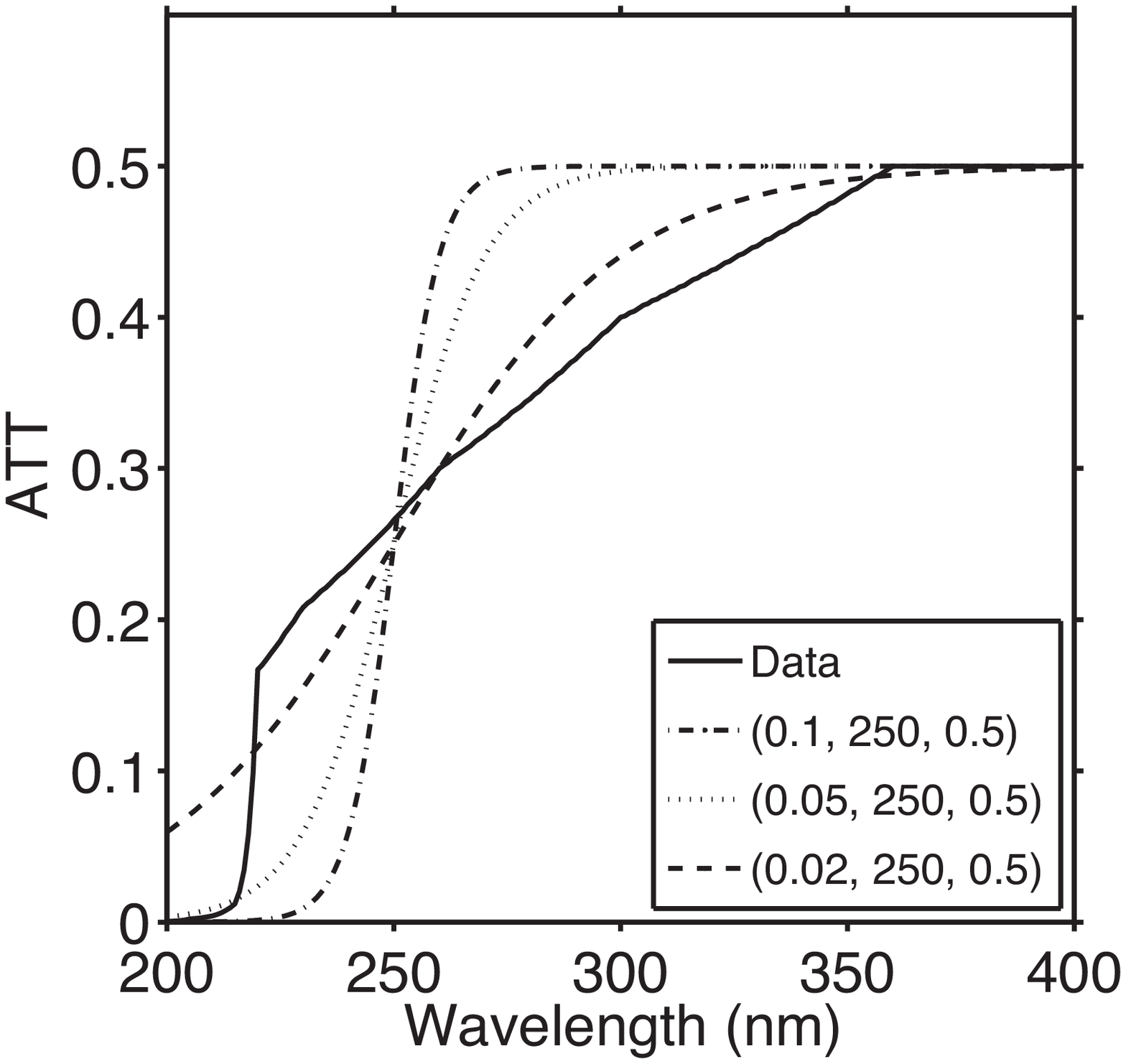,width=0.45\linewidth} \\
\epsfig{file=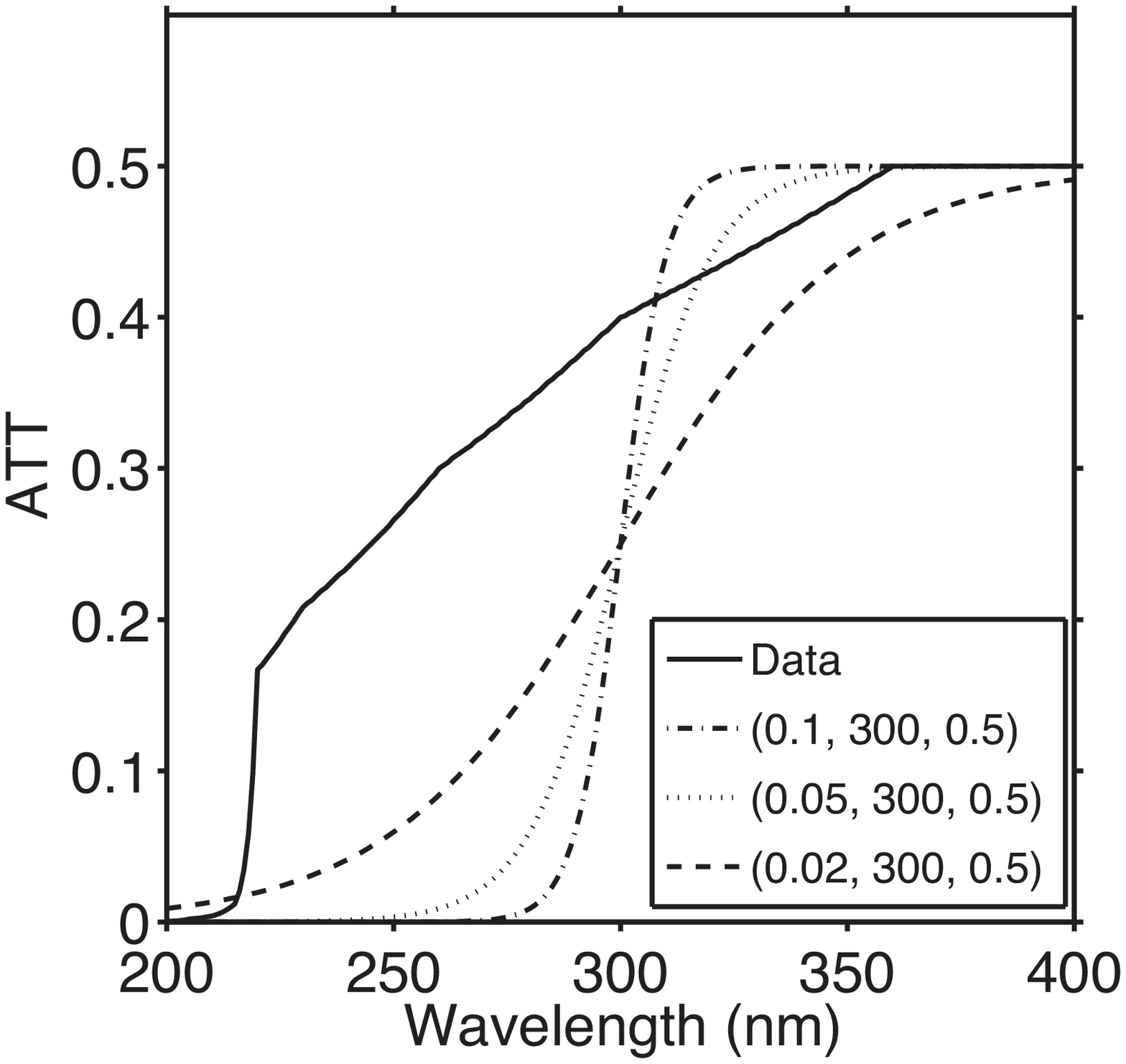,width=0.45\linewidth} \\
\end{tabular}
\caption{
Examples of parameterized attention functions ATT defined through
$(A, B, C)$ (see Eq.~3).  The solid line indicates the proposed case of Earth
3.5~Gyr ago \citep{cock02}.  Note that the attention functions are not intended
to fit the Earth-based data with the tentative exception of $(0.02, 250, 0.5)$.
}
\end{figure*}

\clearpage


\begin{figure*}
\centering
\begin{tabular}{c}
\epsfig{file=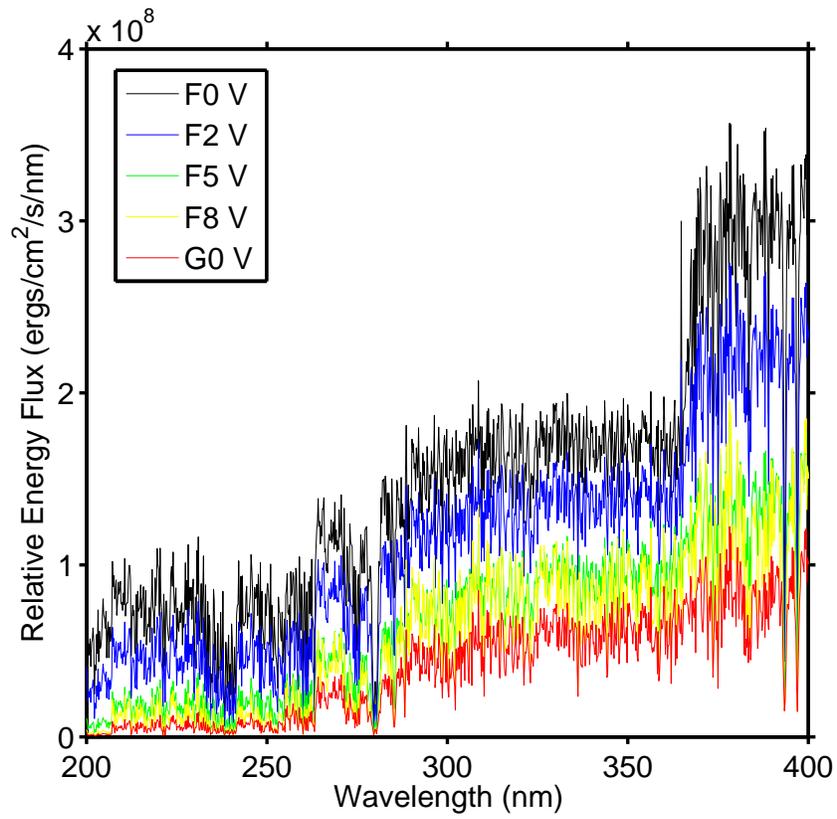,width=0.75\linewidth}
\end{tabular}
\caption{
Stellar spectra as calculated by the PHOENIX code given by \cite{haus92}
and subsequent work (see text for details).
}
\end{figure*}

\clearpage


\begin{figure*}
\centering
\begin{tabular}{c}
\epsfig{file=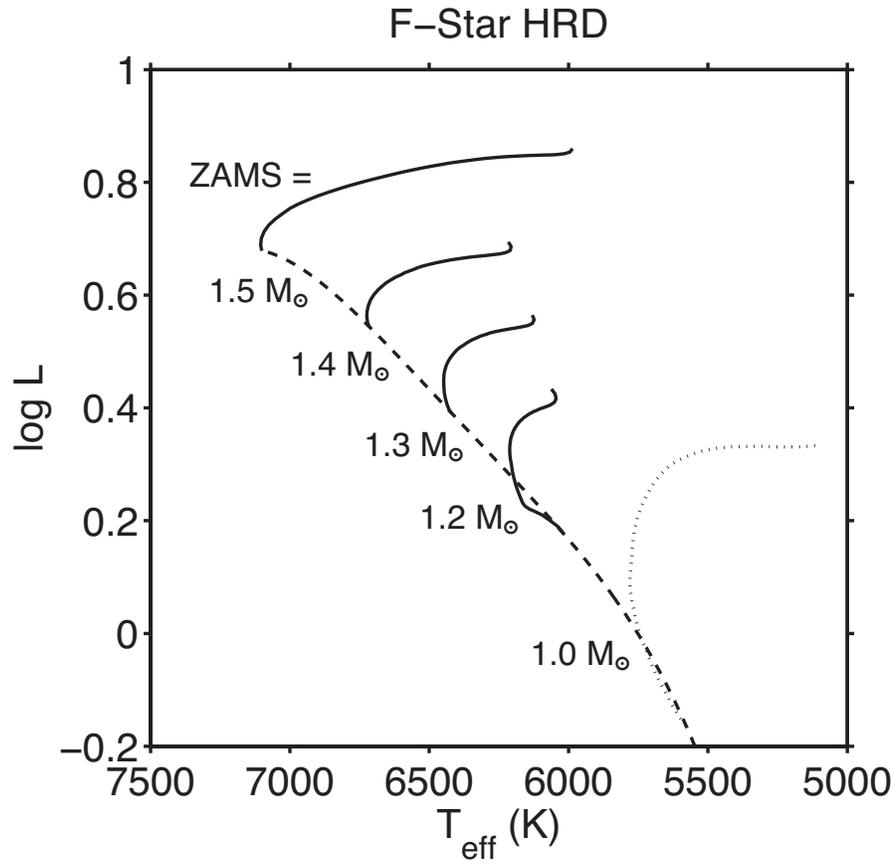,width=0.75\linewidth}
\end{tabular}
\caption{
Stellar evolutionary tracks for a set of F-type stars (solid lines)
denoted by their initial masses with the main-sequence depicted by a
dashed line.  The evolutionary track for the Sun (G2~V) is given as
a dotted line.
}
\end{figure*}

\clearpage


\begin{figure*}
\centering
\begin{tabular}{c}
\epsfig{file=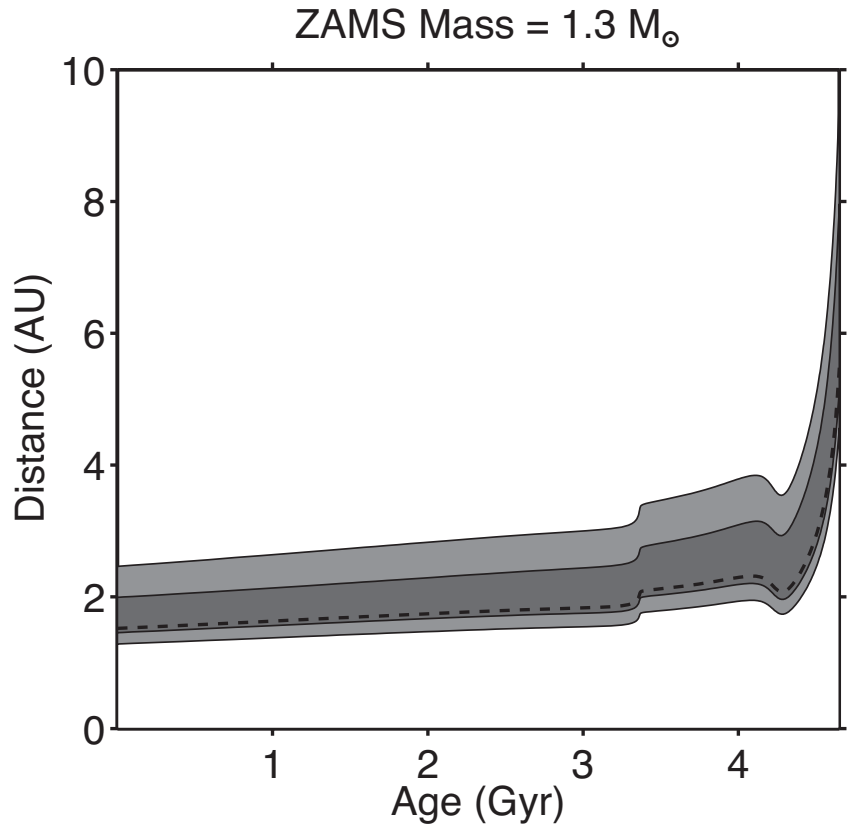,width=0.75\linewidth}
\end{tabular}
\caption{
Evolution of the climatological habitable zone for a ZAMS star with a mass of
1.3~$M_\odot$.  Dark gray color indicates the CHZ, whereas light gray color
indicates the GHZ.  Earth-equivalent positions are depicted by a dashed line.
The temporary increase of the inner and outer limits of the habitable zone,
between the age of 3.5 and 4.2 billion years, reflects an episode of higher
luminosity due to the onset of hydrogen shell burning.
}
\end{figure*}

\clearpage


\begin{figure*}
\centering
\begin{tabular}{cc}
\epsfig{file=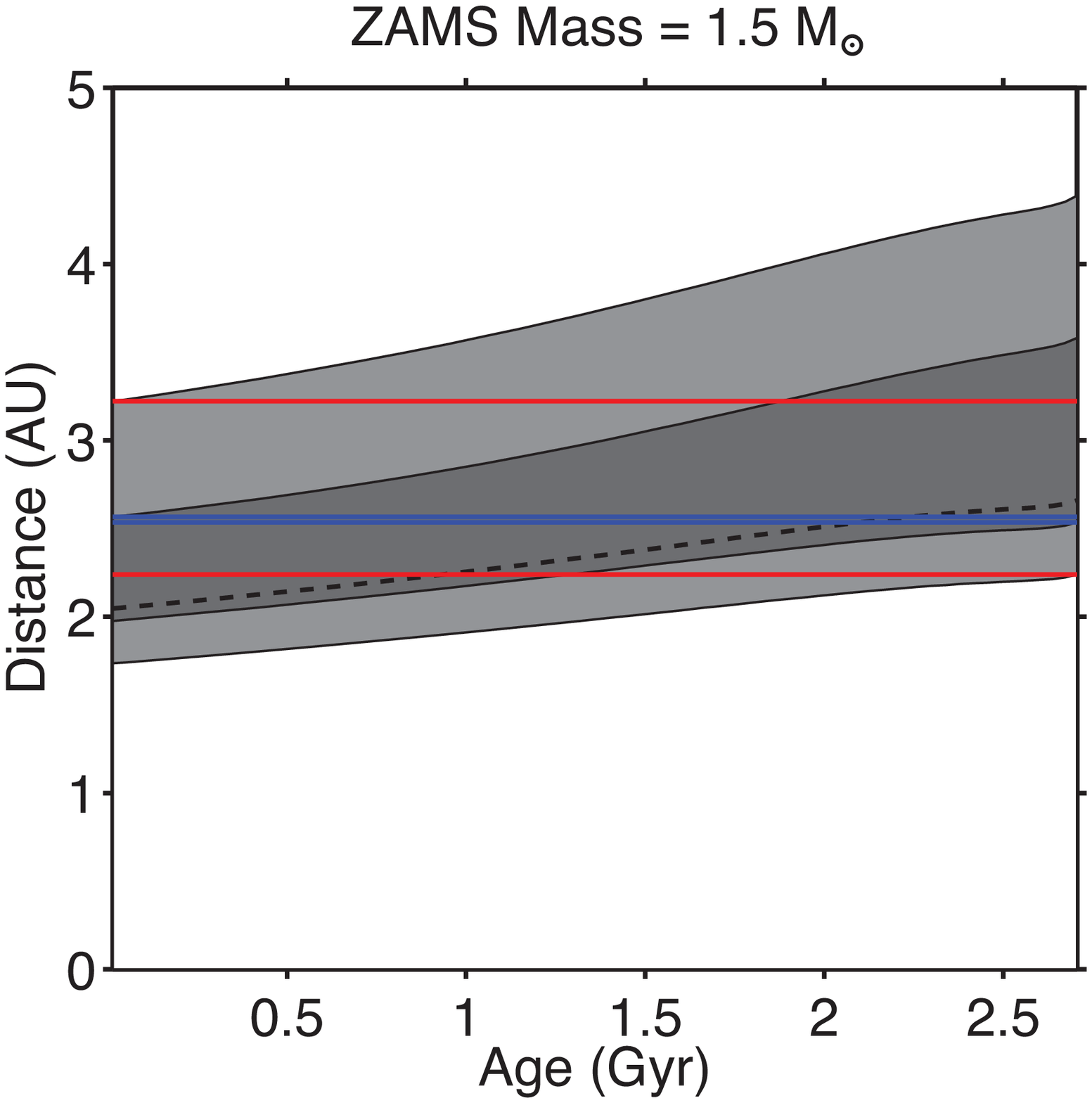,width=0.45\linewidth} &
\epsfig{file=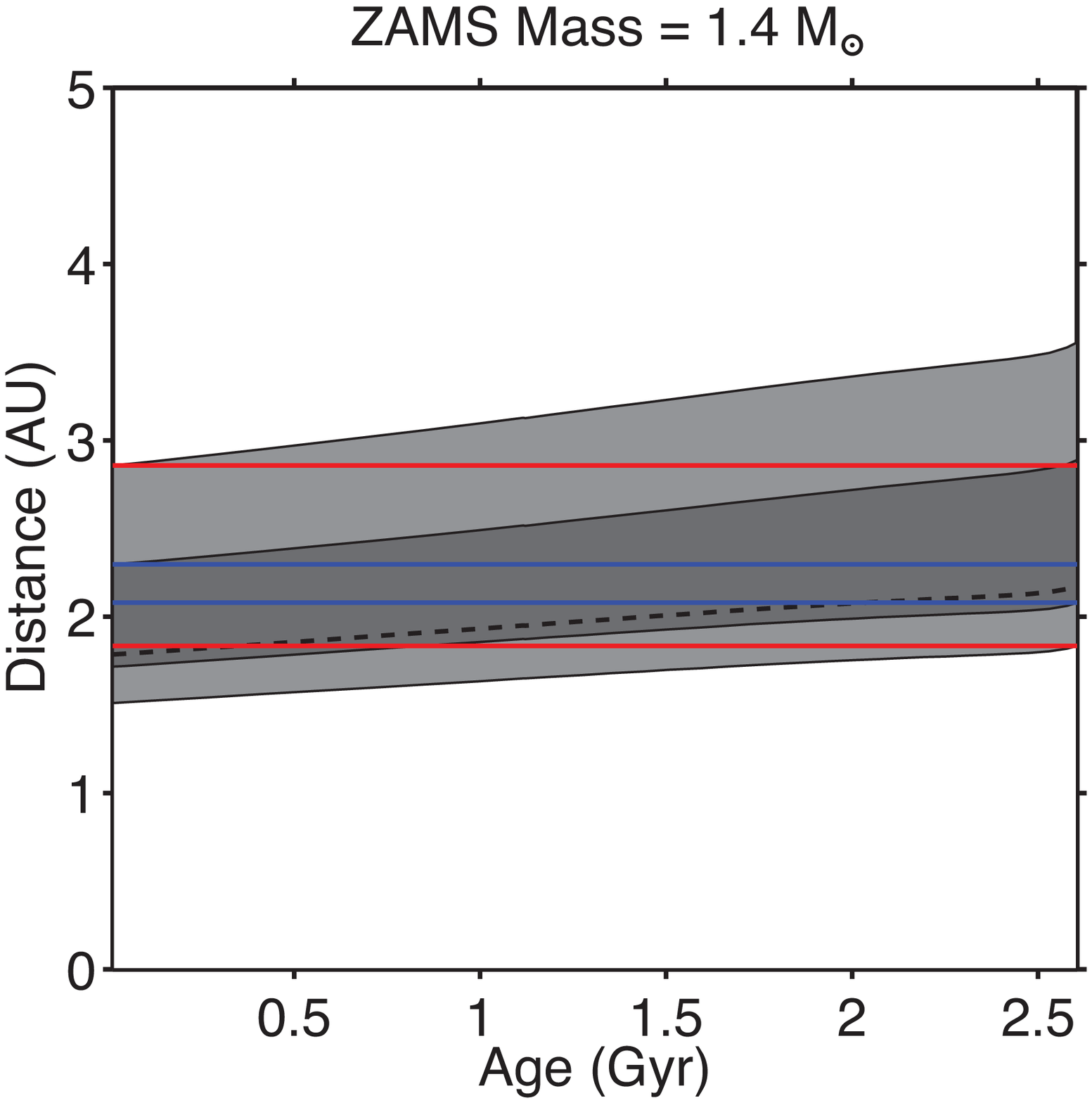,width=0.45\linewidth} \\
\epsfig{file=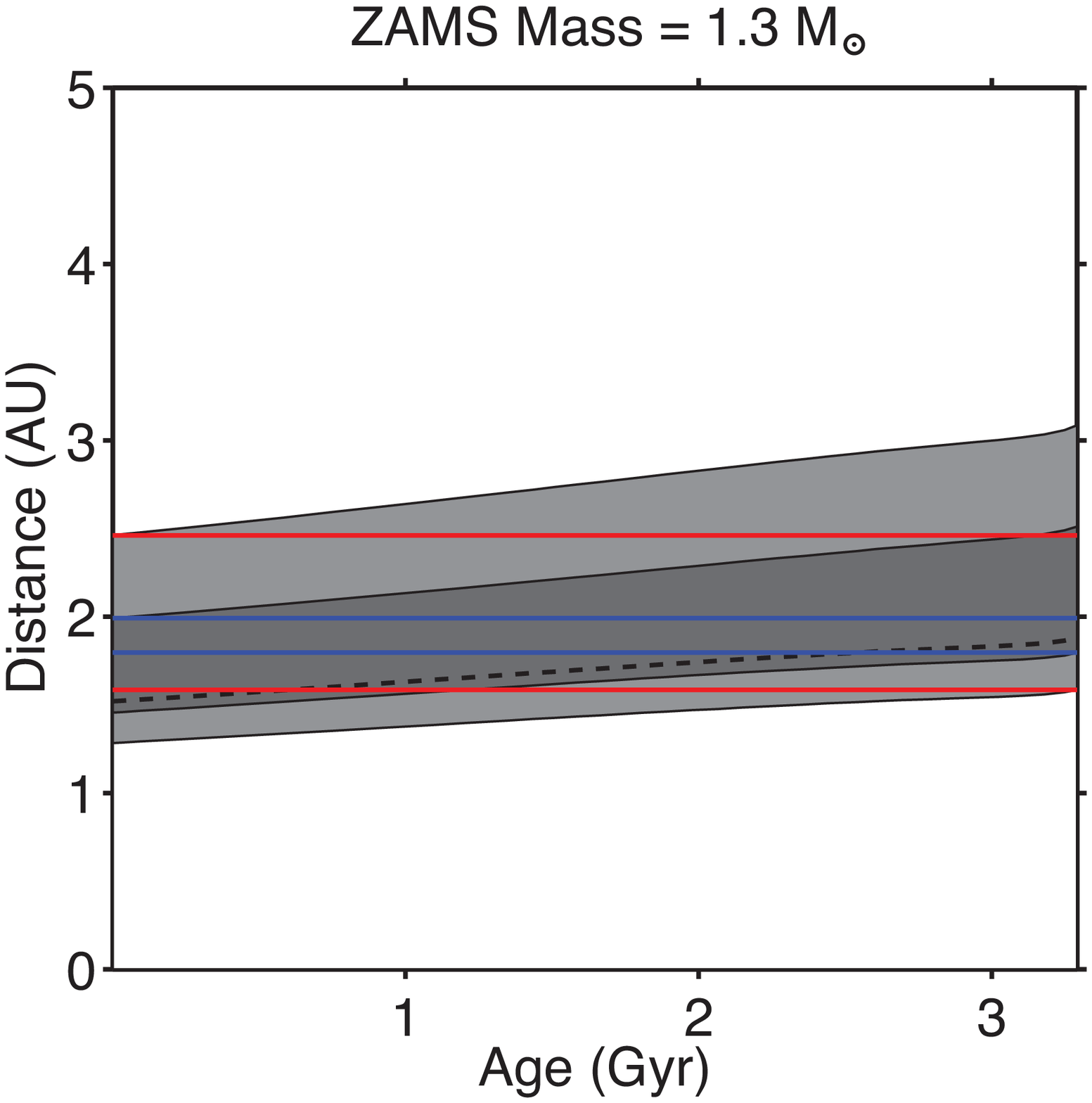,width=0.45\linewidth} &
\epsfig{file=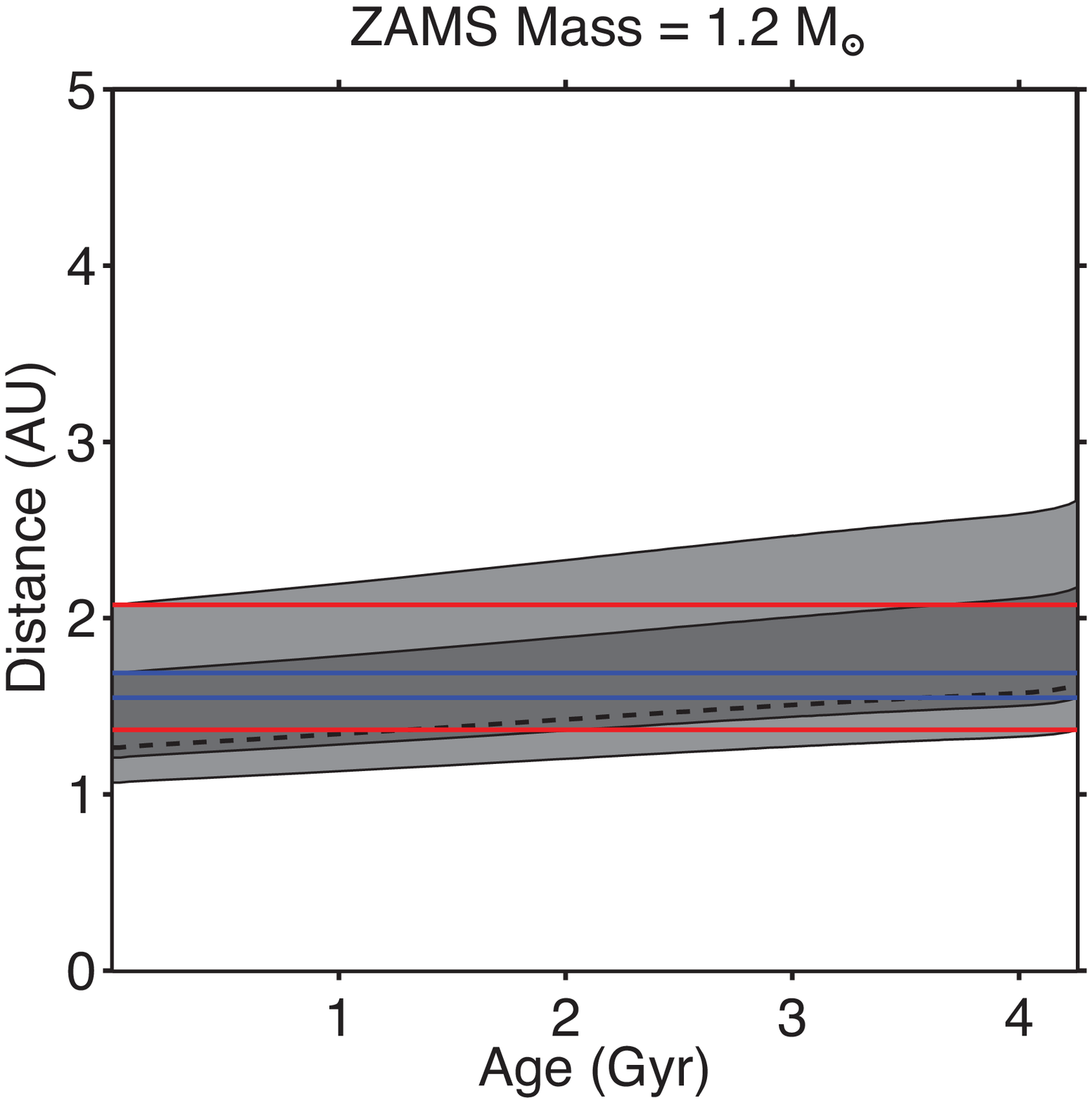,width=0.45\linewidth} \\
\end{tabular}
\caption{
Magnified segments of the evolution of CHZs (dark gray) and GHZs
(light gray) for selected F-type stars (indicated by their stellar masses)
during their lifetimes on the main-sequence.  The corresponding
Earth-equivalent positions, which evolve with time, are depicted by dashed lines.
The horizontal blue lines indicate the inner and outer limits of the continuous
domains of the CHZs, whereas the horizontal red lines indicate the inner and outer
limits of the continuous domains of the GHZs.  Note that in the main text, we
introduce an average (i.e., time-independent) Earth-equivalent position located
within the CHZs obtained through interpolation between the conservative outer limit
at the ZAMS and the conservative inner limit at the end of main-sequence evolution.
}
\end{figure*}

\clearpage


\begin{figure*}
\centering
\begin{tabular}{cc}
\epsfig{file=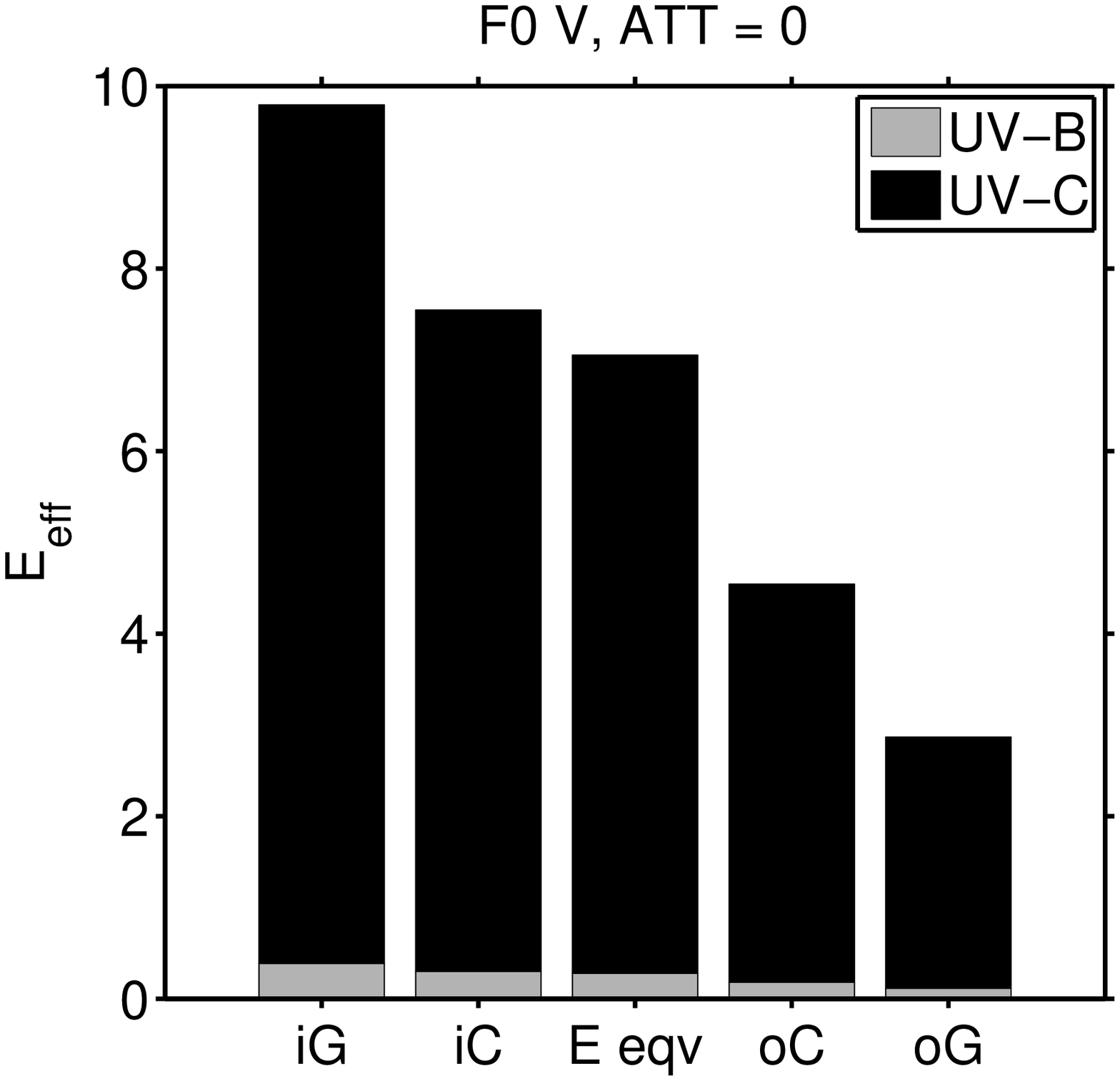,width=0.45\linewidth} &
\epsfig{file=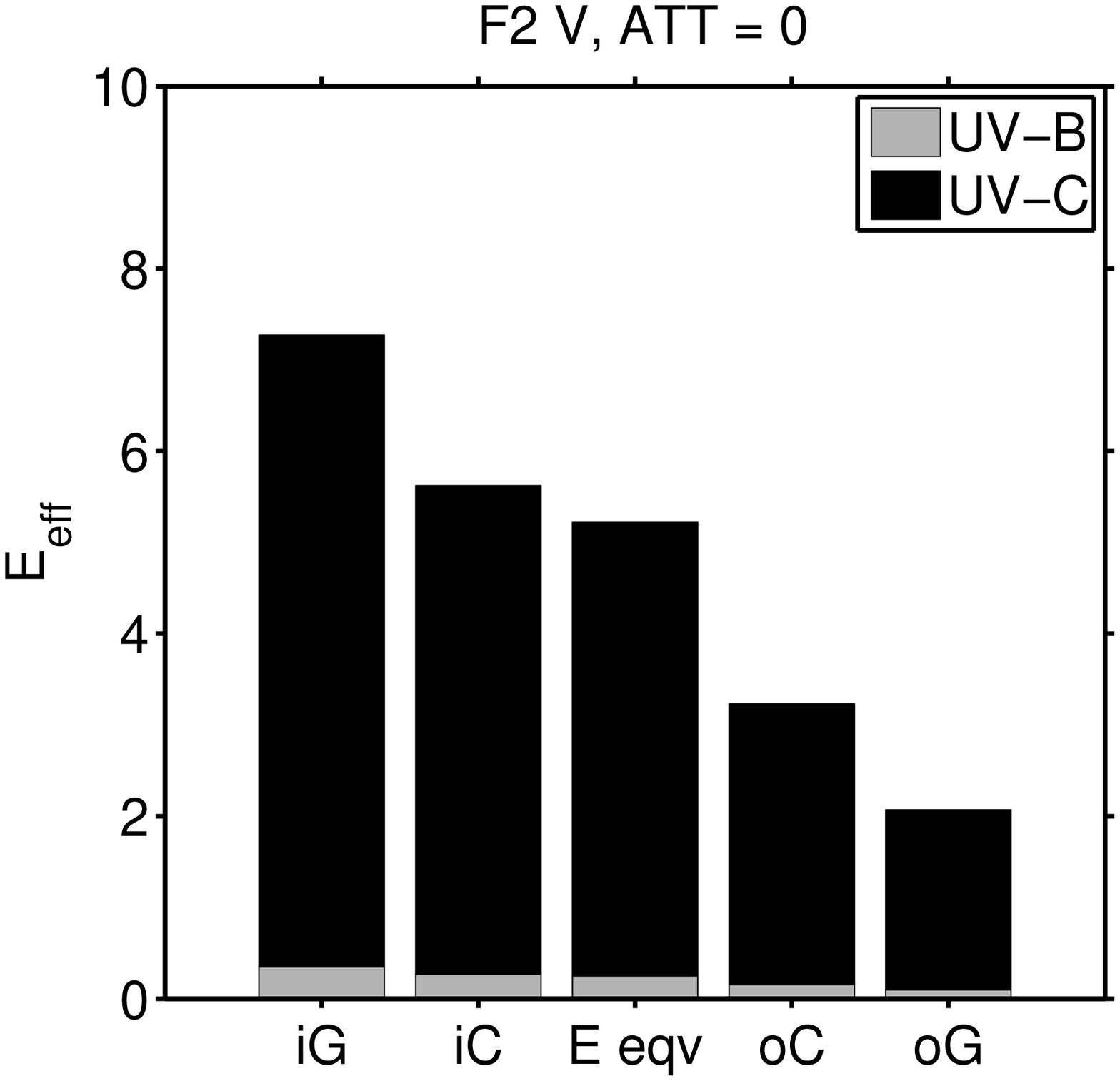,width=0.45\linewidth} \\
\epsfig{file=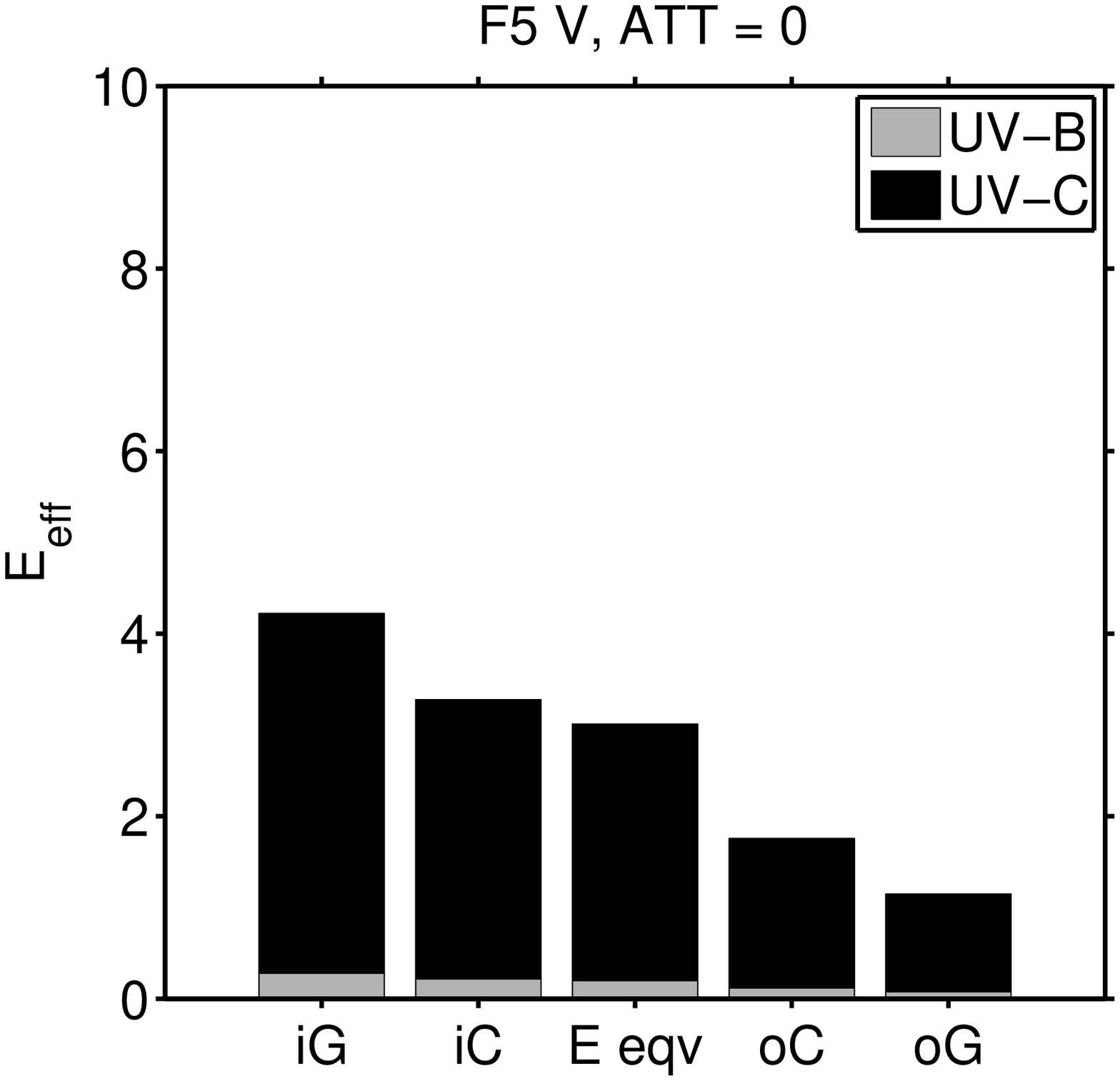,width=0.45\linewidth} &
\epsfig{file=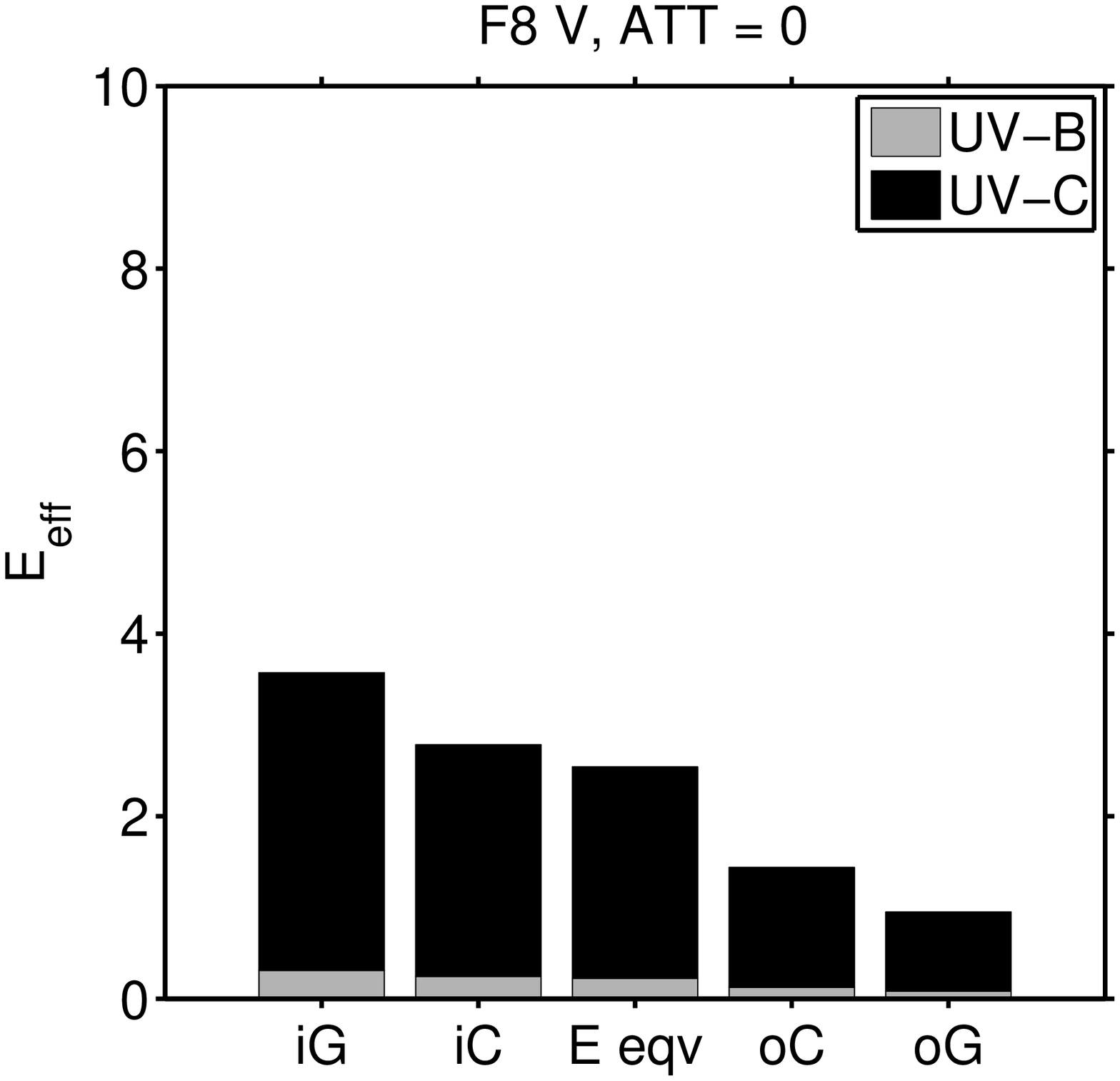,width=0.45\linewidth} \\
\end{tabular}
\caption{
Damage due to UV-B and UV-C inflicted upon DNA (taken as a representative
biomolecule) in the envionments of F0~V, F2~V, F5~V, and F8~V stars determined
at the inner limits of the GHZs and CHZs (iG, iC) and
at the outer limits of the GHZs and CHZs (oG, oC),
as well as for Earth-equivalent positions.  Most of the damage occurs through
UV-C, whereas minor damage occurs through UV-B.  Damage associated with UV-A
is unidentifyable and has therefore been omitted (see text).
No planetary attenuation is taken into account (ATT=0).
}
\end{figure*}

\clearpage


\begin{figure*}
\centering
\begin{tabular}{cc}
\epsfig{file=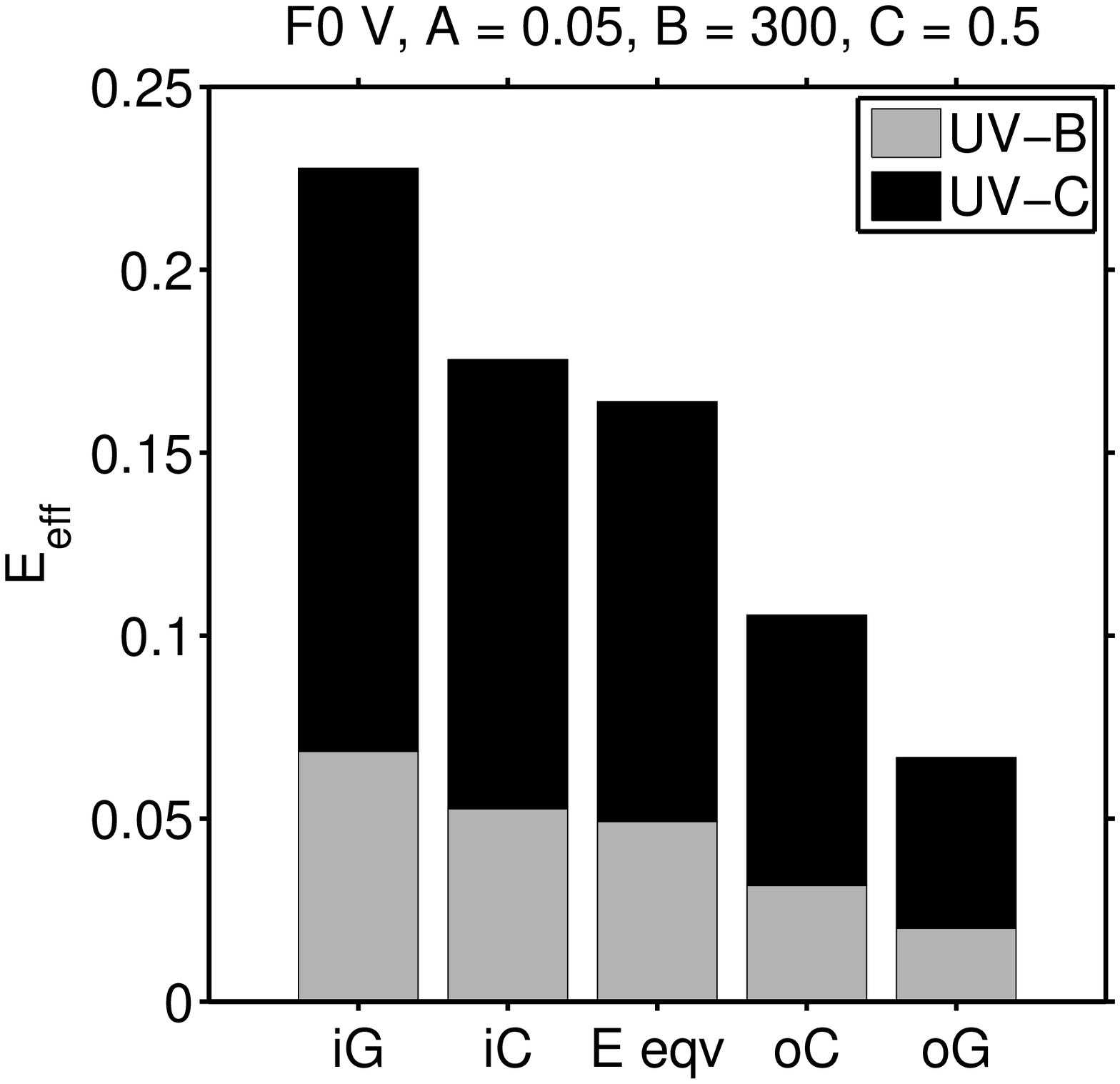,width=0.45\linewidth} &
\epsfig{file=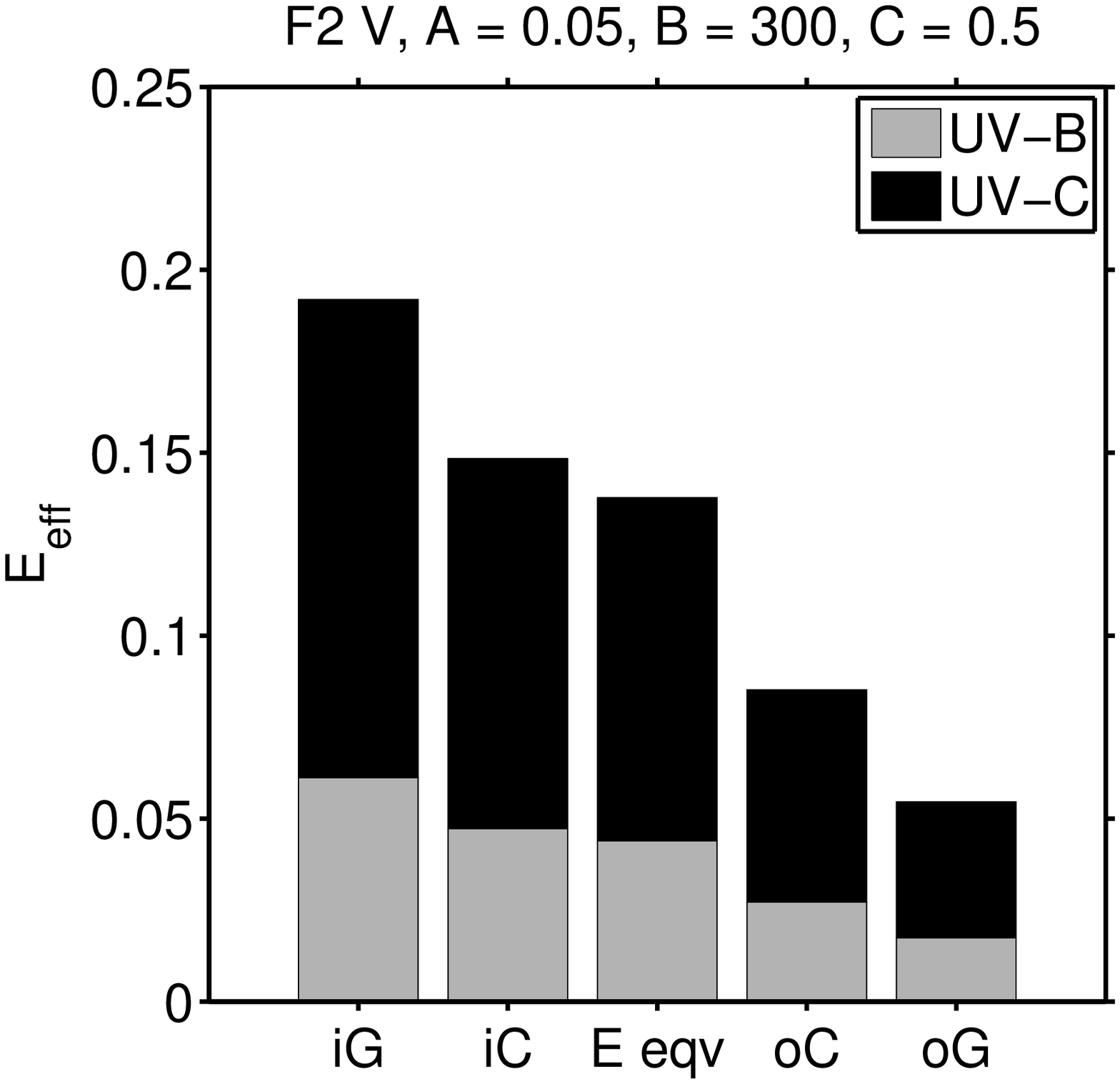,width=0.45\linewidth} \\
\epsfig{file=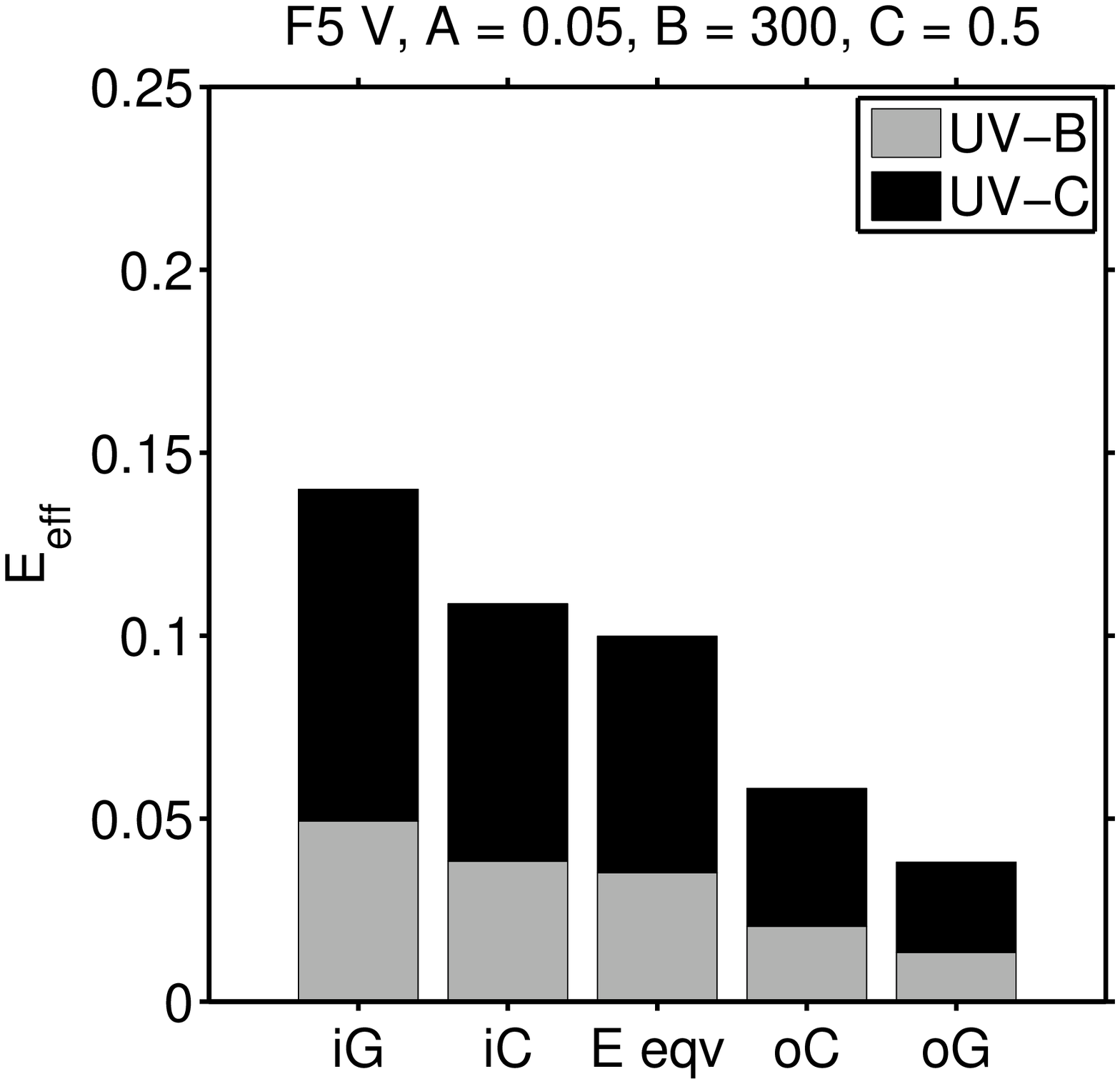,width=0.45\linewidth} &
\epsfig{file=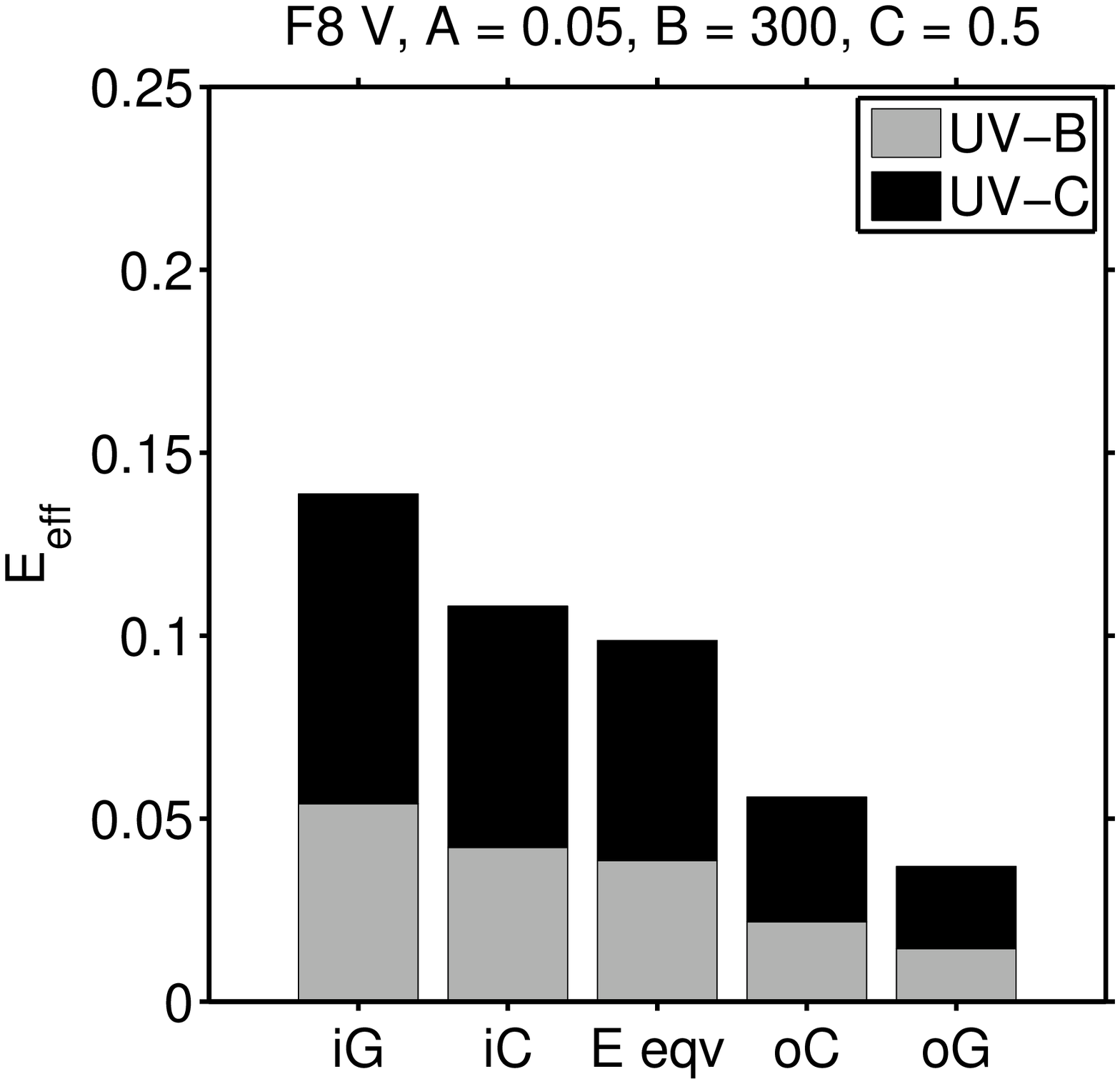,width=0.45\linewidth} \\
\end{tabular}
\caption{
Same as Fig.~8, but with inclusion of planetary attenuation.
The attenuation parameters $A$, $B$, and $C$ are specified (see Eq.~3).
}
\end{figure*}

\clearpage


\begin{figure*}
\centering
\begin{tabular}{cc}
\epsfig{file=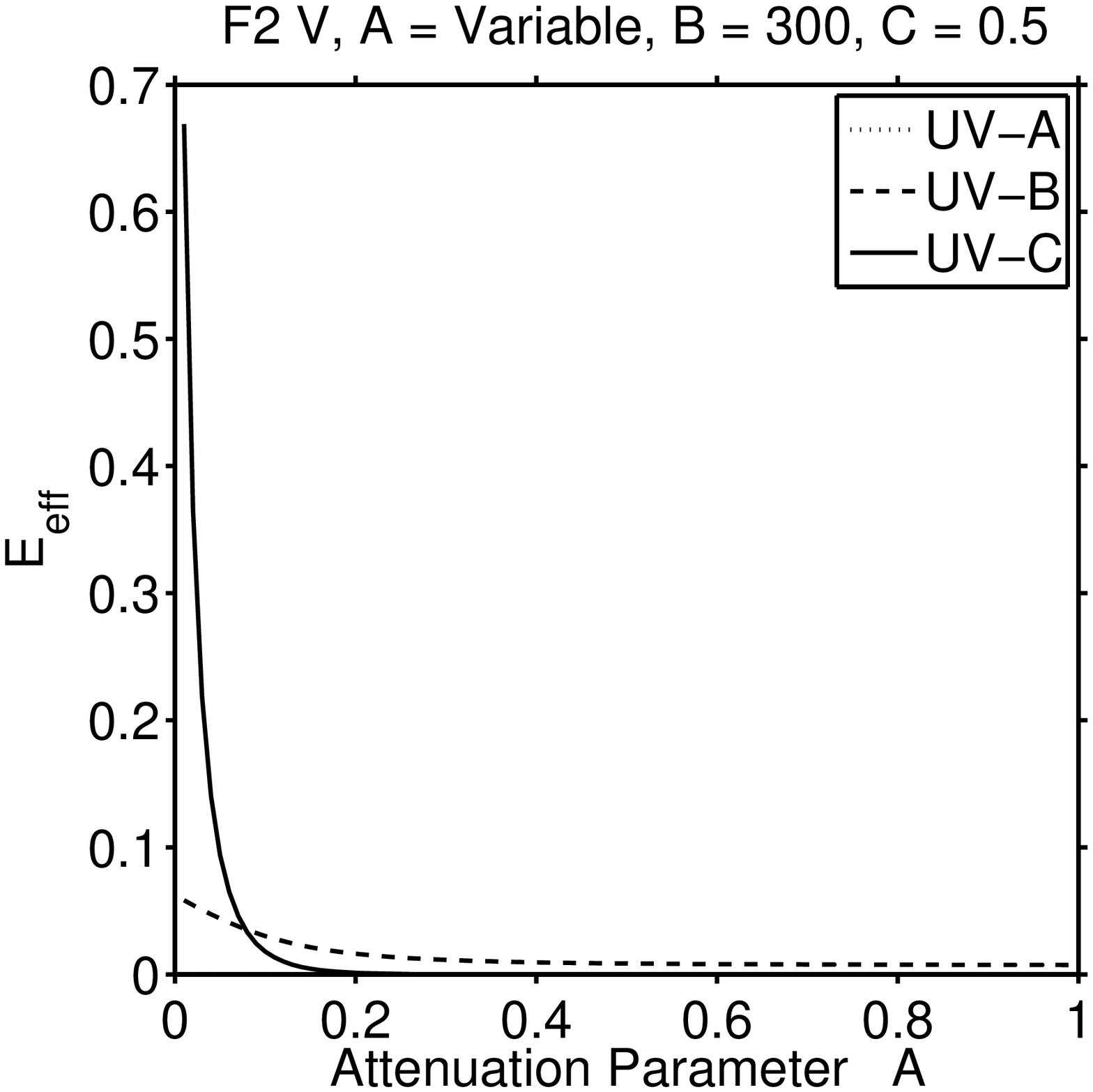 ,width=0.45\linewidth} &
\epsfig{file=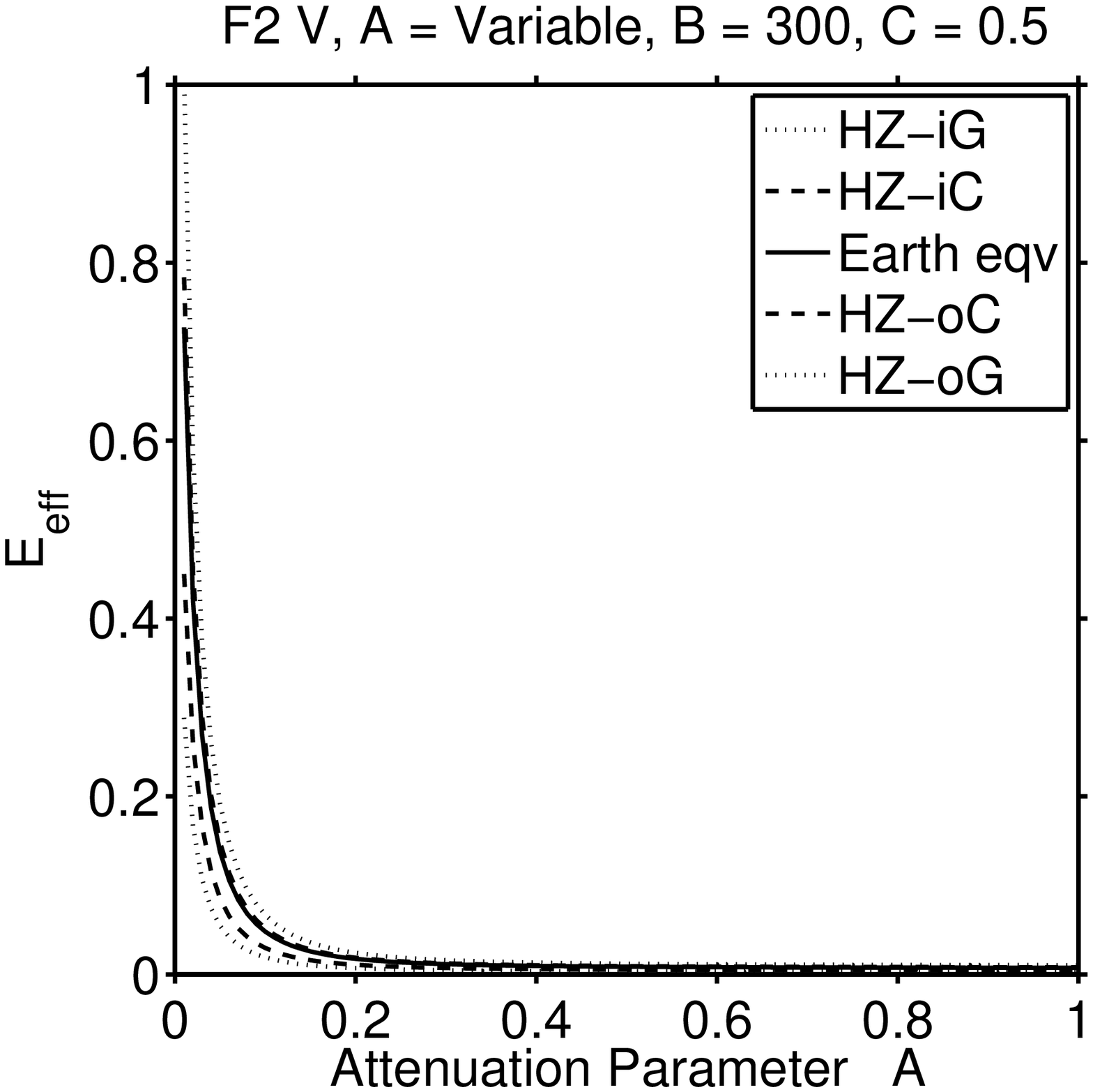,width=0.45\linewidth} \\
\epsfig{file=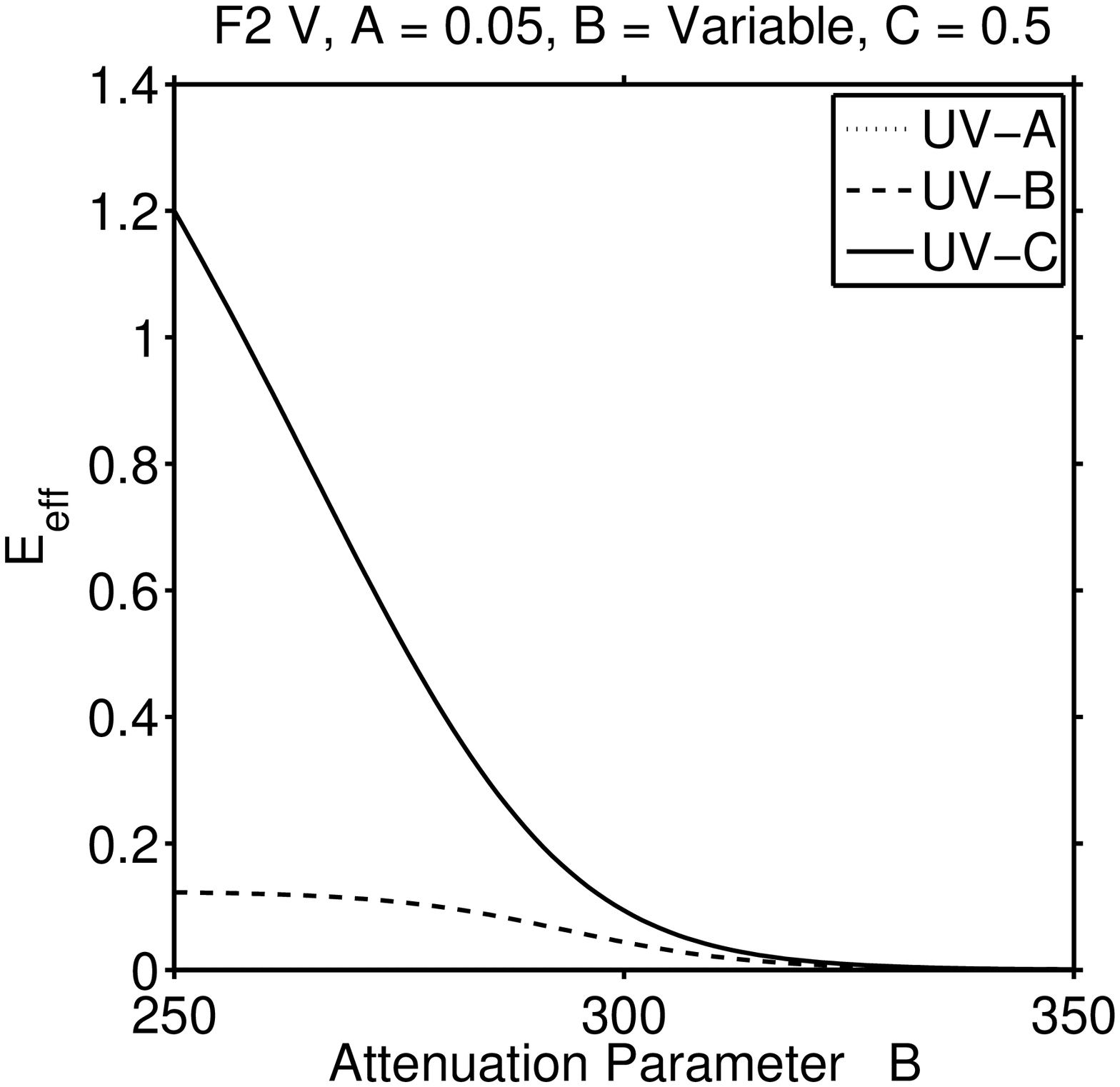 ,width=0.45\linewidth} &
\epsfig{file=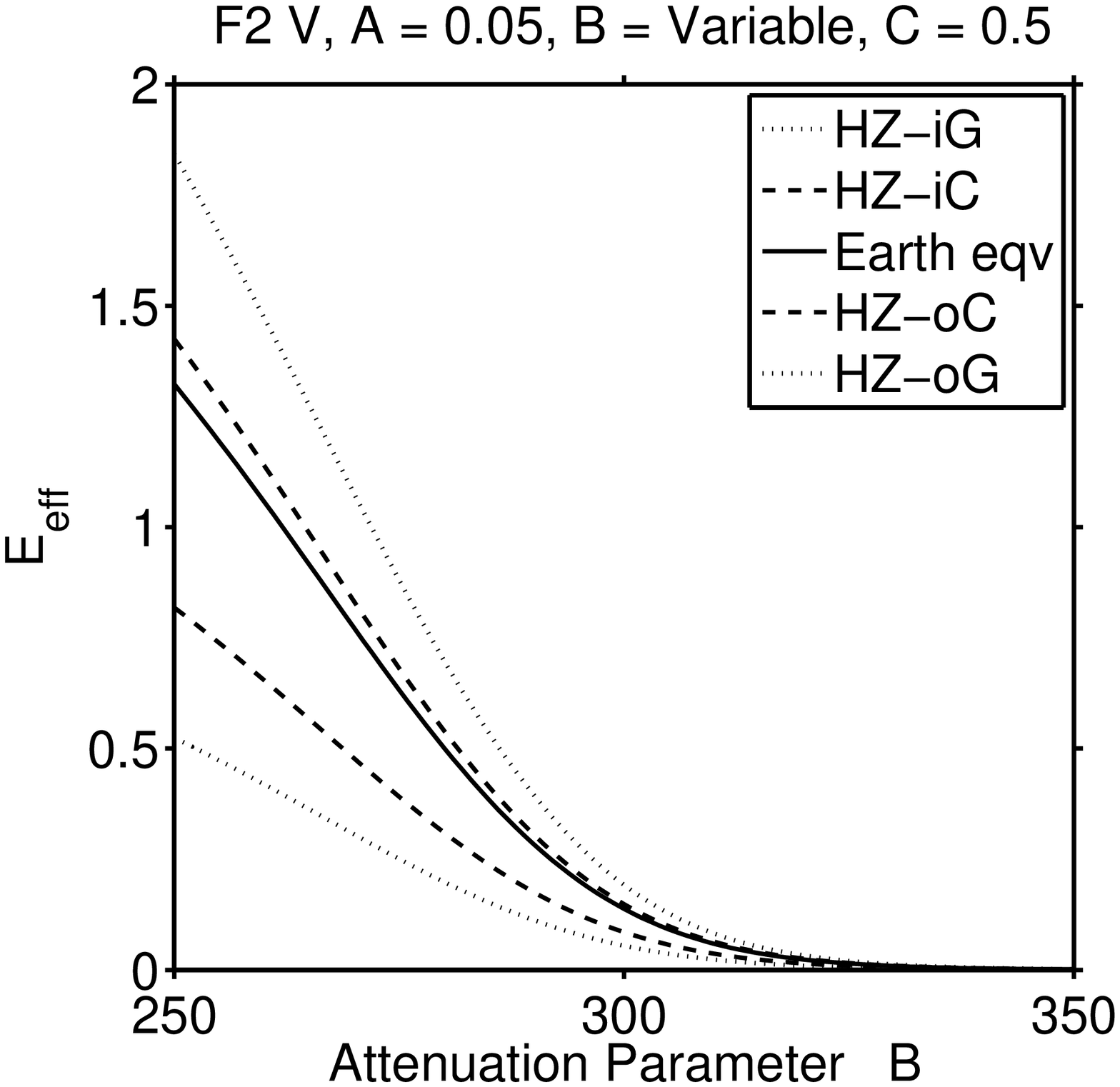,width=0.45\linewidth} \\
\end{tabular}
\caption{
Damage inflicted upon DNA at specified positions within the
habitable zone for an F2~V star.  Planetary atmospheric attenuation
is taken into account with parameters as specified.
}
\end{figure*}

\clearpage


\begin{figure*}
\centering
\begin{tabular}{cc}
\epsfig{file=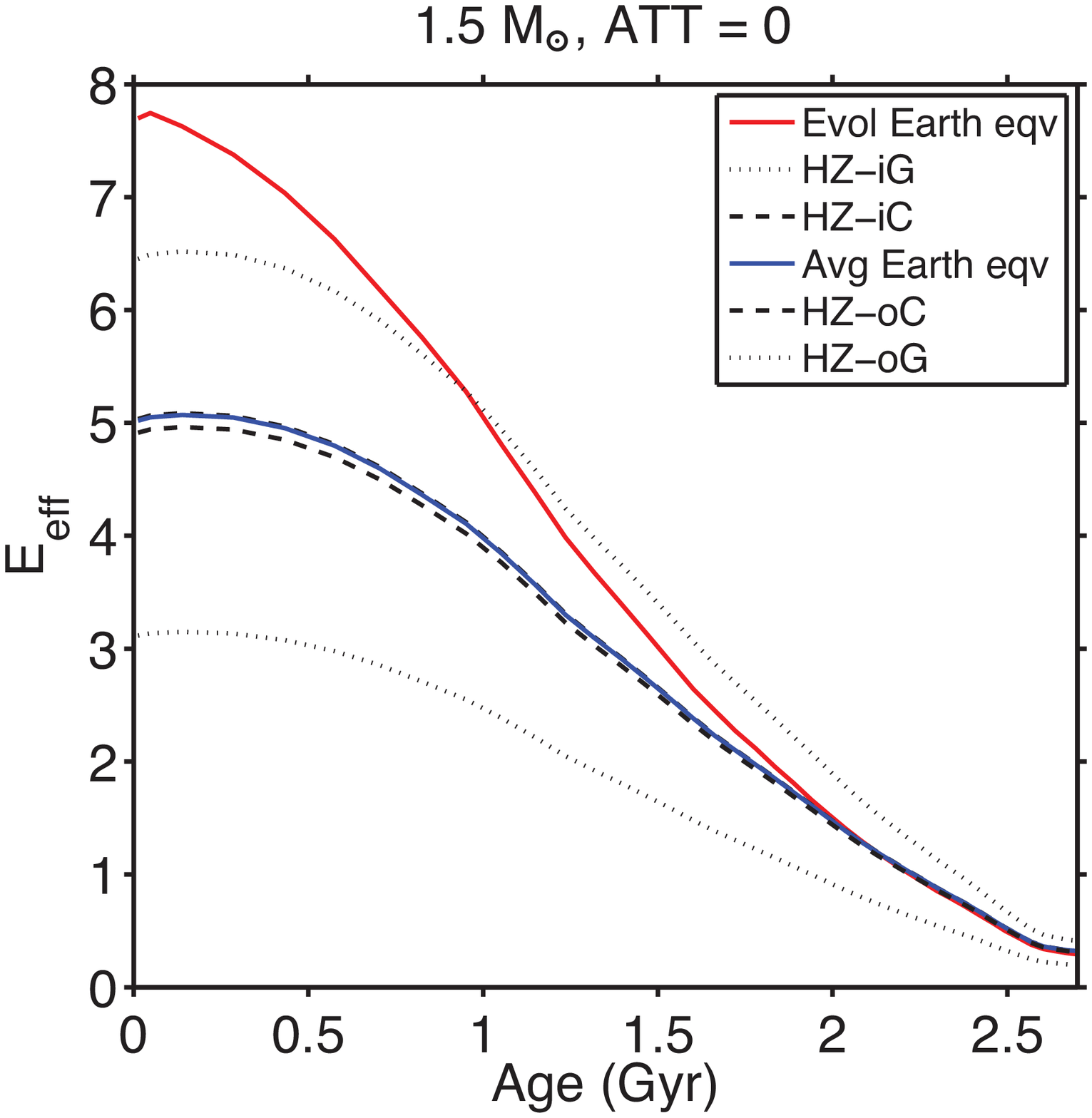,width=0.45\linewidth} &
\epsfig{file=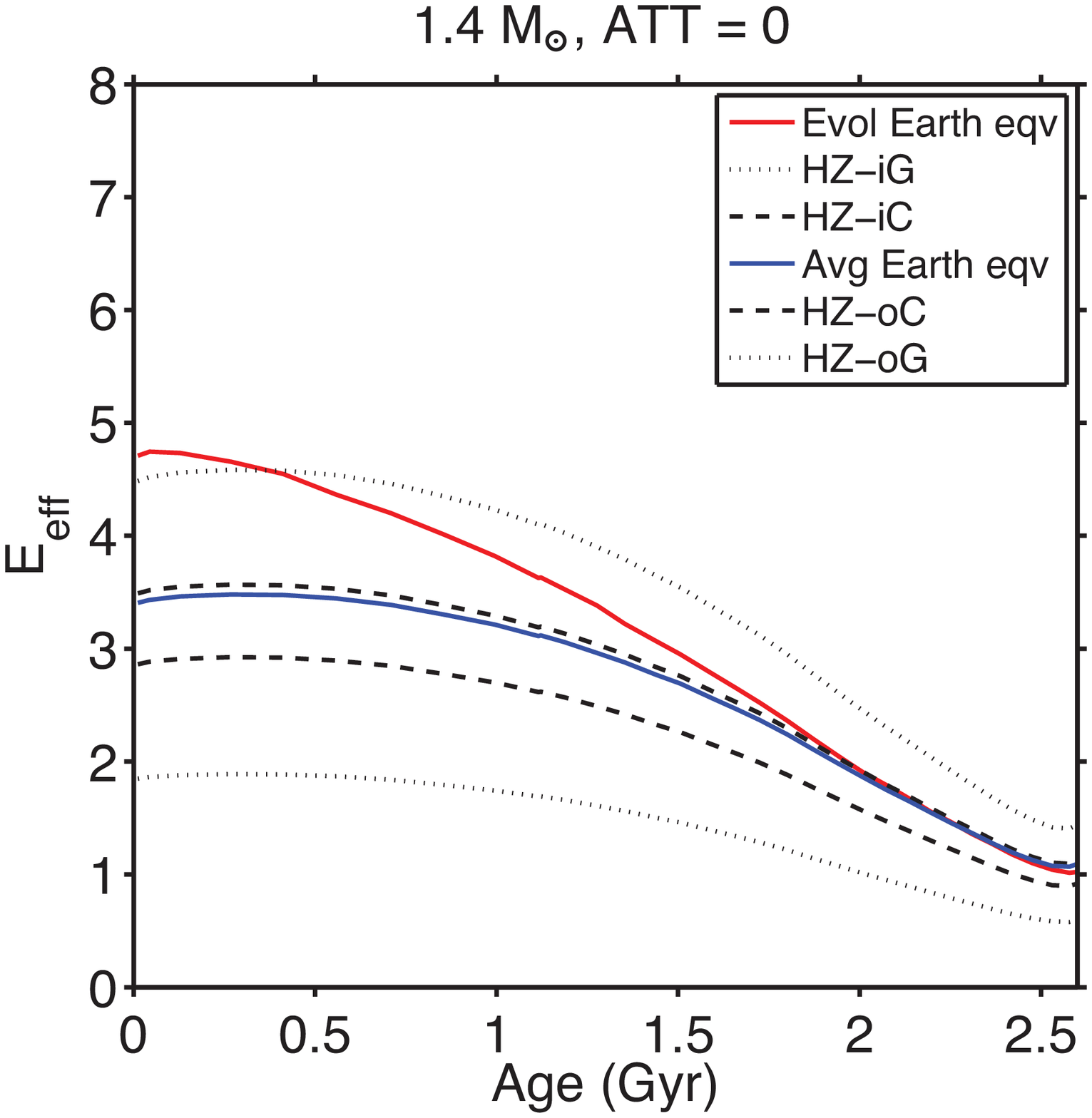,width=0.45\linewidth} \\
\epsfig{file=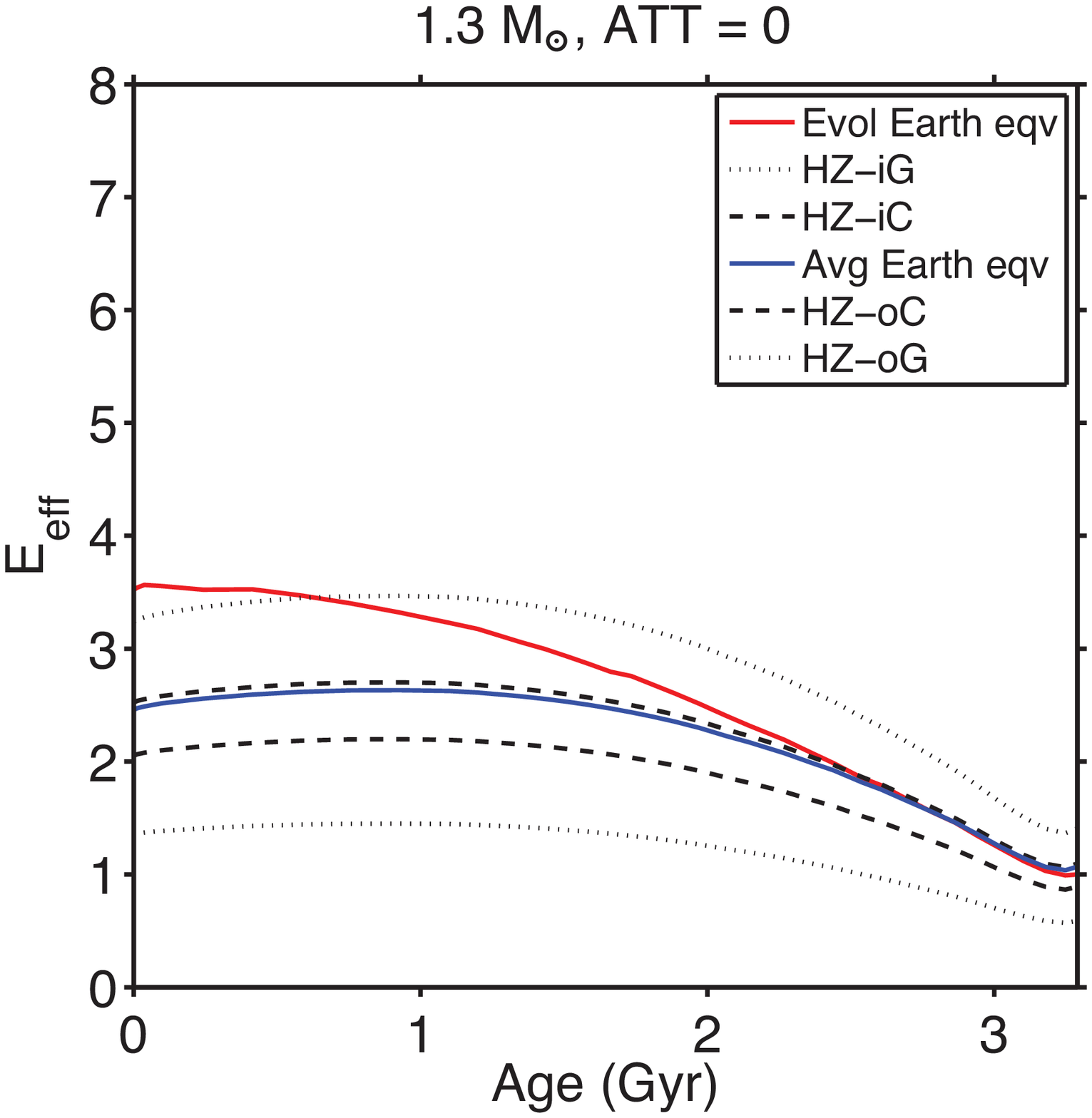,width=0.45\linewidth} &
\epsfig{file=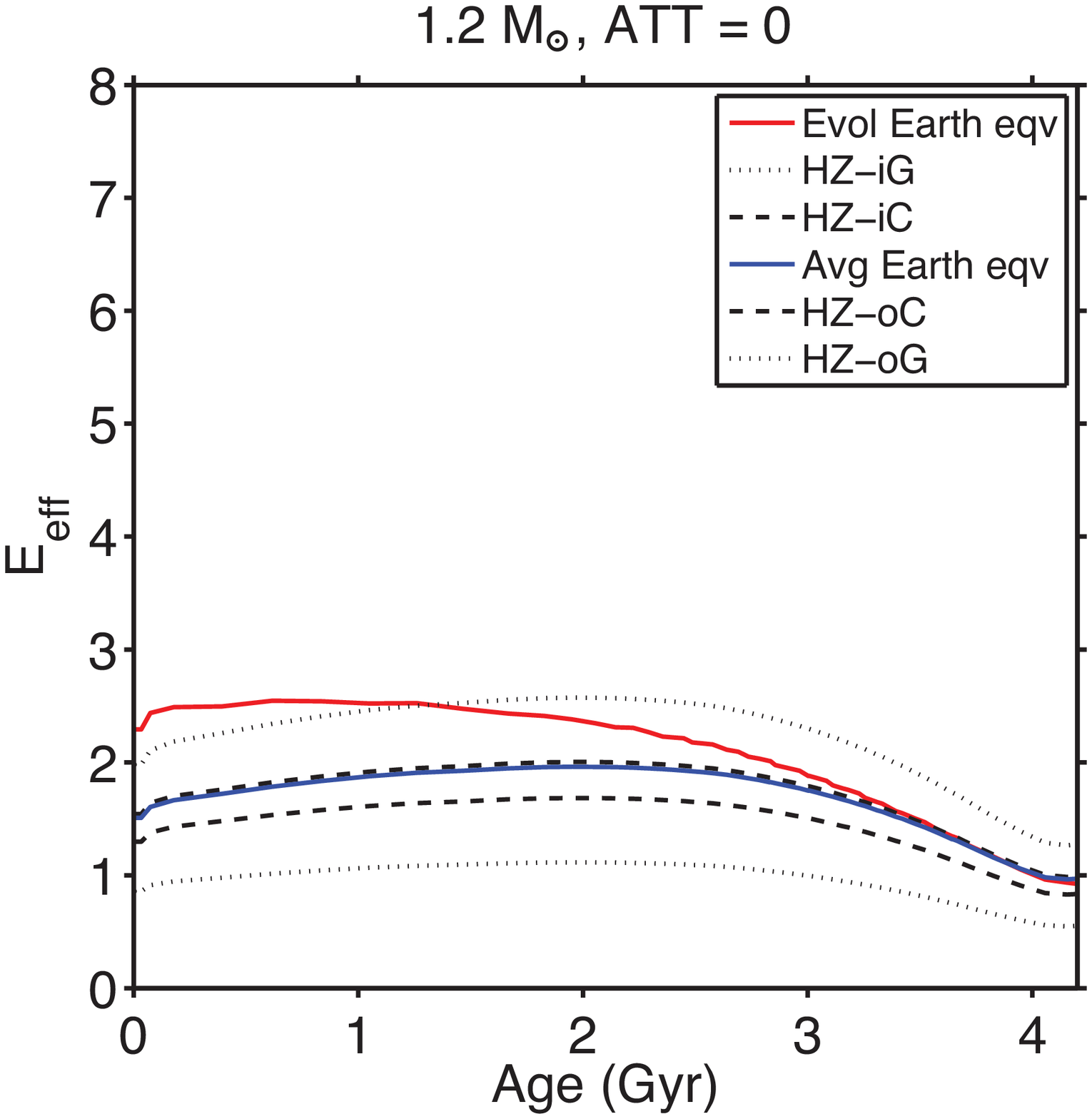,width=0.45\linewidth} \\
\end{tabular}
\caption{
Damage inflicted upon DNA at specified positions within the
habitable zone of various F-type stars, characterized by their masses,
at different ages.  No planetary attenuation is taken into account
(ATT=0).  Results are given for the HZ-iG, HZ-iC, HZ-oC, and HZ-oG
in reference to the continuous domains of the CHZs and GHZs, respectively
(see Fig.~7 and Table 4).  The behaviors of the E$_{\rm eff}$ are largely
due to the changes of the stellar luminosities.  We also show the results
for the average Earth-equivalent positions (embedded into the continuous
domains of the CHZs) and the evolving Earth-equivalent positions.  The
latter conform to the dashed lines of Fig.~7.
}
\end{figure*}

\clearpage


\begin{figure*}
\centering
\begin{tabular}{cc}
\epsfig{file=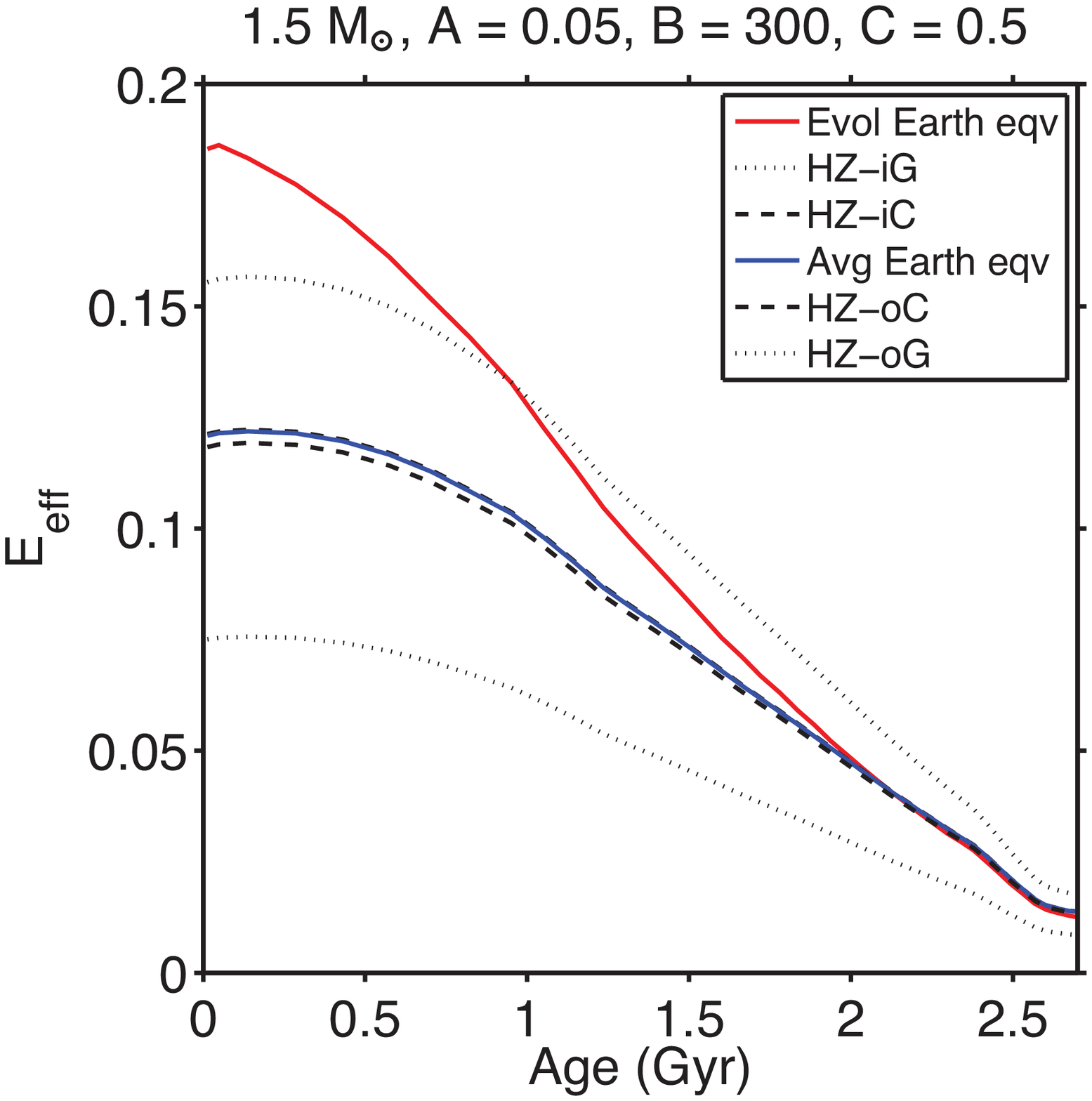,width=0.45\linewidth} &
\epsfig{file=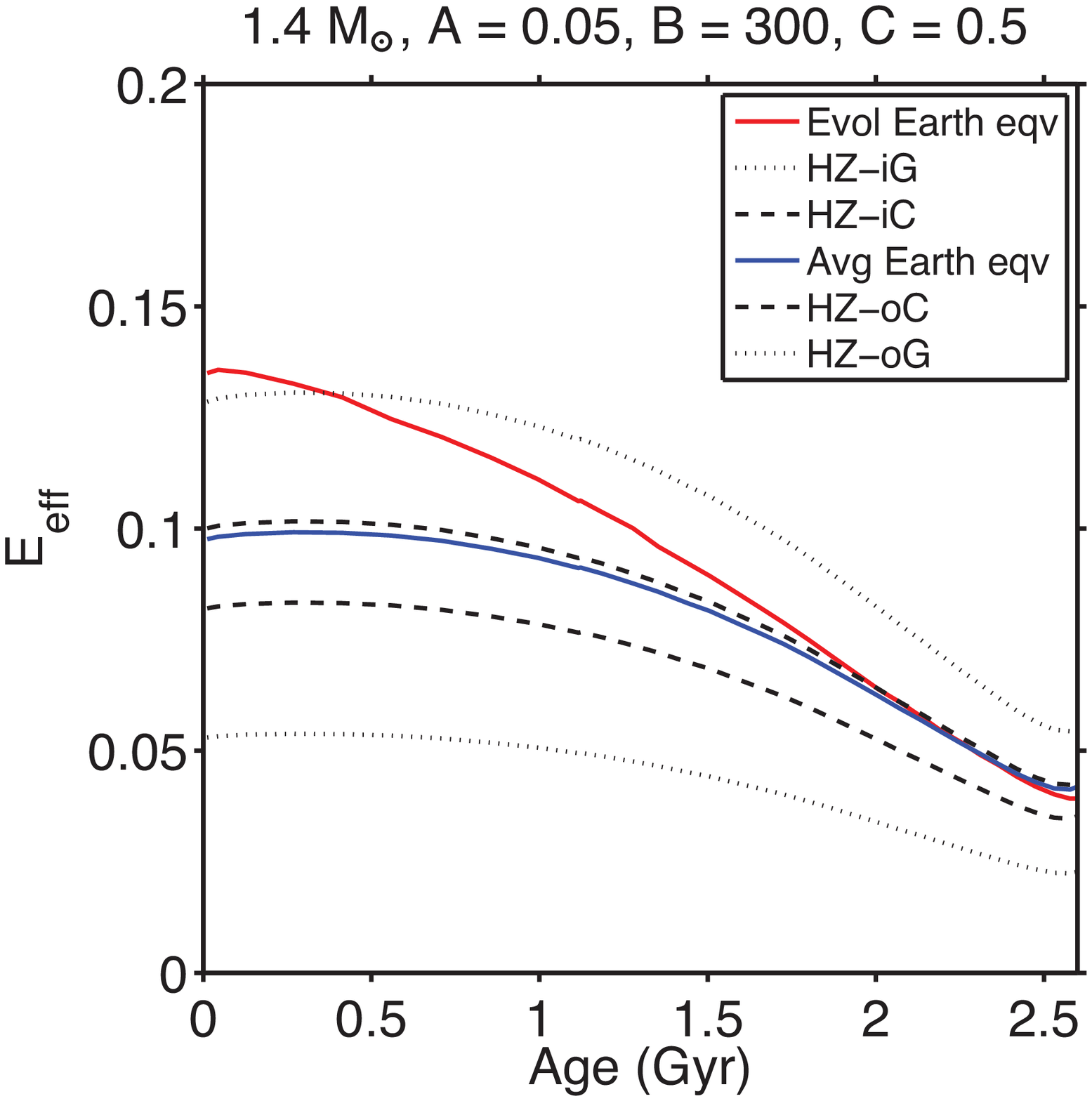,width=0.45\linewidth} \\
\epsfig{file=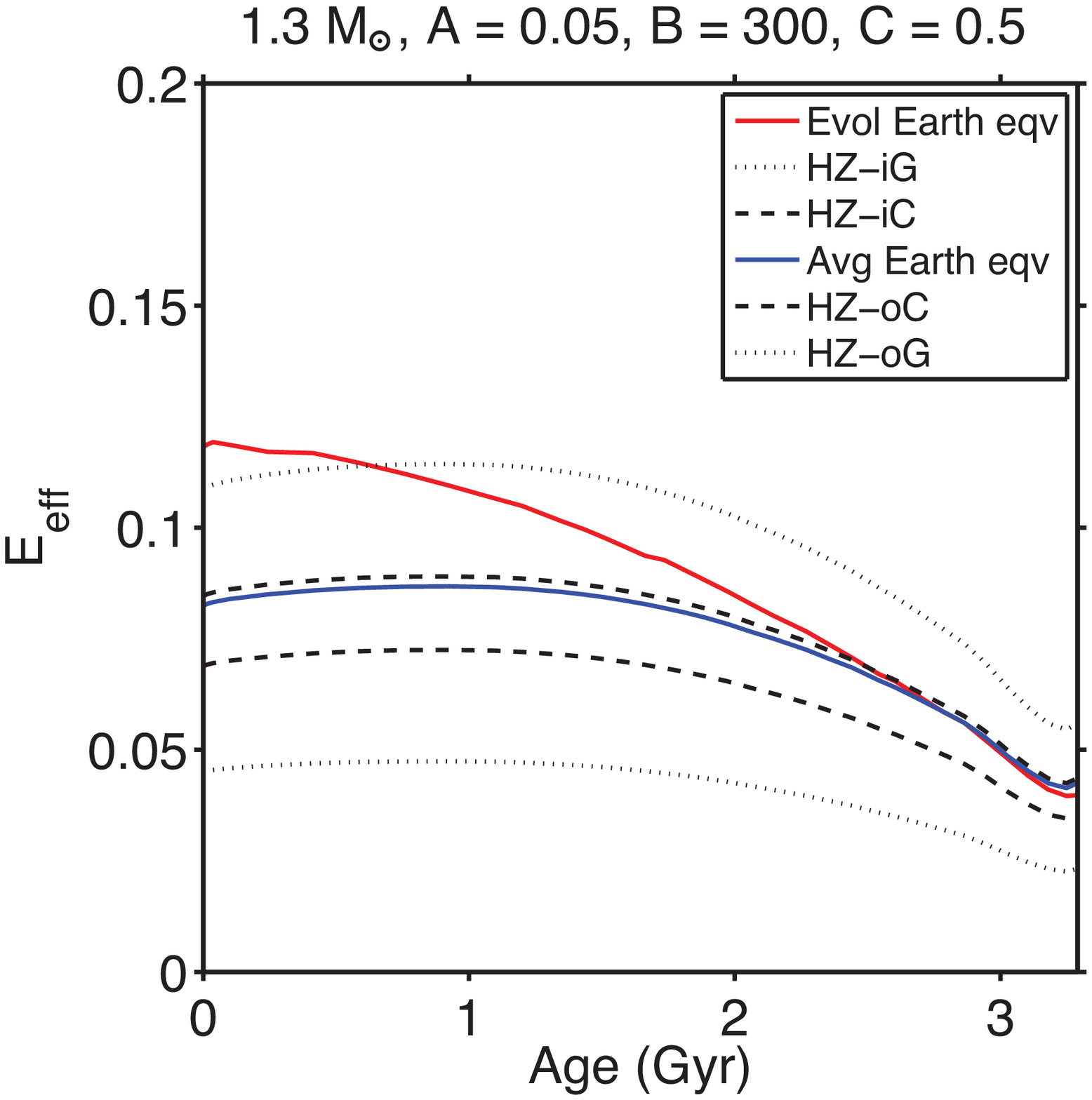,width=0.45\linewidth} &
\epsfig{file=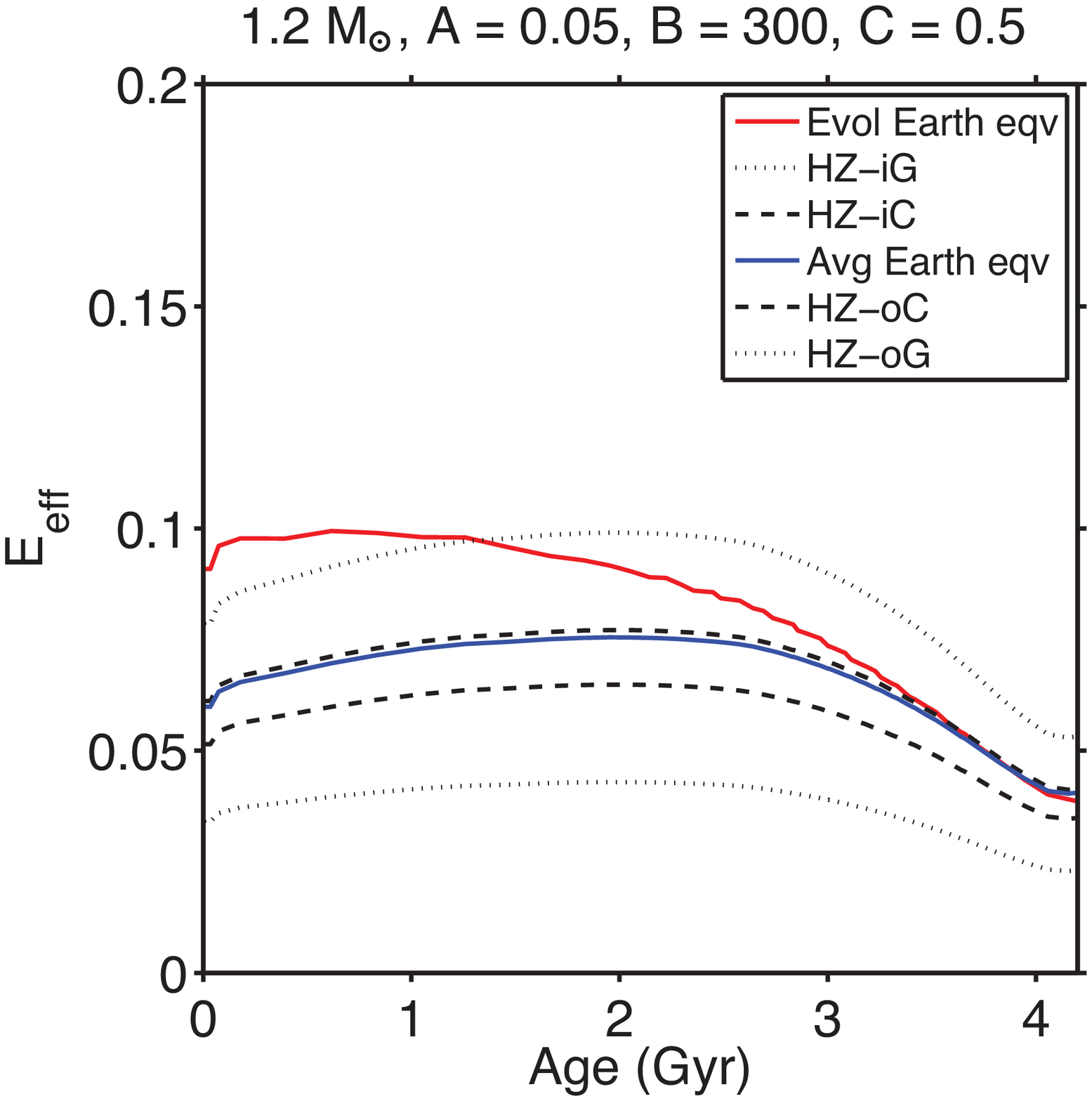,width=0.45\linewidth} \\
\end{tabular}
\caption{
Damage inflicted upon DNA at specified positions within the
habitable zone of various F-type stars, characterized by their masses,
at different ages.  Planetary atmospheric attenuation is taken into
account with parameters as specified.  See Fig.~11 for additional
information.
}
\end{figure*}

\clearpage



\begin{deluxetable}{lcccc}
\tablecaption{F-star reference parameters}
\tablewidth{0pt}
\tablehead{
Spectral Type   & $T_{\rm eff}$ & $R_\ast$    & $L_\ast$    & $M_\ast$    \\
...             & (K)           & ($R_\odot$) & ($L_\odot$) & ($M_\odot$)
}
\startdata
    F0  &  7200  &  1.62  &  6.33  &  1.60 \\
    F2  &  6890  &  1.48  &  4.43  &  1.50 \\
    F5  &  6440  &  1.40  &  3.03  &  1.25 \\
    F8  &  6200  &  1.20  &  1.91  &  1.10 \\
    G0  &  6050  &  1.12  &  1.51  &  1.05 \\ 
\enddata
\end{deluxetable}

\clearpage


\thispagestyle{empty}
\begin{landscape}
\begin{deluxetable}{lcccccccccccc}
\tablecaption{F-star parameters during stellar main-sequence evolution}
\tablewidth{0pt}
\tablehead{
Age   & $T_{\rm eff}$ & $L_\ast$    & Spectral Type & $T_{\rm eff}$ & $L_\ast$    & Spectral Type & $T_{\rm eff}$ & $L_\ast$    & Spectral Type & $T_{\rm eff}$ & $L_\ast$    & Spectral Type \\
(Gyr) & (K)           & ($L_\odot$) & ...           & (K)           & ($L_\odot$) & ...           & (K)           & ($L_\odot$) & ...           & (K)           & ($L_\odot$) & ...     
}
\startdata
\omit & \multicolumn{3}{c}{$M = 1.2~M_\odot$}  & \multicolumn{3}{c}{$M = 1.3~M_\odot$}  & \multicolumn{3}{c}{$M = 1.4~M_\odot$}  & \multicolumn{3}{c}{$M = 1.5~M_\odot$}  \\
\noalign{\smallskip}
\hline
\noalign{\smallskip}
 0.5  &  6178         &  1.770   &  F8 $-22$~K & 6444         &  2.644   &  F5          &  6720         &  3.780   &  F3 $+20$~K & 7084         &  5.228   &  F1 $+84$~K  \\
 1.0  &  6195         &  1.878   &  F8         & 6447         &  2.839   &  F5          &  6682         &  4.077   &  F3 $-18$~K & 6983         &  5.728   &  F1 $-17$~K  \\
 1.5  &  6205         &  1.996   &  F8         & 6432         &  3.036   &  F5          &  6589         &  4.361   &  F4 $+59$~K & 6775         &  6.247   &  F3 $+75$~K  \\
 2.0  &  6211         &  2.121   &  F8 $+11$~K & 6386         &  3.224   &  F6 $+26$~K  &  6428         &  4.591   &  F5 $-12$~K & 6482         &  6.753   &  F5 $+42$~K  \\
 2.5  &  6204         &  2.247   &  F8         & 6303         &  3.383   &  F7 $+23$~K  &  6218         &  4.755   &  F8 $+18$~K & 6117         &  7.047   &  F9          \\
 3.0  &  6178         &  2.366   &  F8 $-22$~K & 6174         &  3.493   &  F8 $-26$~K  &  ...          &  ...     &  ...        & ...          &  ...     &  ...         \\
 3.5  &  6130         &  2.468   &  F9         & ...          &  ...     & ...          &  ...          &  ...     &  ...        & ...          &  ...     &  ...         \\
 4.0  &  6058         &  2.555   &  G0         & ...          &  ...     & ...          &  ...          &  ...     &  ...        & ...          &  ...     &  ...         \\
\enddata
\tablecomments{
For example, an expression such as F8 $-22$~K means that the stellar effective temperature is 22~K less than
that for a standard F8 main-sequence star; for parameters see \cite{haus99a} and subsequent work.  Deviations
from stellar spectral types are omitted if the difference is found to be less than 10~K.
}
\end{deluxetable}
\end{landscape}

\clearpage



\begin{deluxetable}{lccccc}
\tablecaption{Stellar habitable zones including stellar main-sequence evolution ---
extreme~positions}
\tablewidth{0pt}
\tablehead{
$M_\ast$     & $t_{\rm ms}$   &  iG    &  iC    &  oC    &  oG  \\
($M_\odot$)  & (Gyr)          &  (AU)  &  (AU)  &  (AU)  & (AU)
}
\startdata
 1.2  &   4.255  &  1.066  &  1.208  &  2.174  &  2.668     \\
 1.3  &   3.291  &  1.283  &  1.455  &  2.511  &  3.085     \\
 1.4  &   2.605  &  1.510  &  1.716  &  2.888  &  3.555     \\
 1.5  &   2.706  &  1.735  &  1.975  &  3.581  &  4.388     \\
\enddata
\end{deluxetable}

\clearpage



\begin{deluxetable}{lccccc}
\tablecaption{Stellar habitable zones including stellar main-sequence evolution --- continuous~domains}
\tablewidth{0pt}
\tablehead{
$M_\ast$     & $t_{\rm ms}$   &  iG    &  iC    &  oC    &  oG  \\
($M_\odot$)  & (Gyr)          &  (AU)  &  (AU)  &  (AU)  & (AU)
}
\startdata
 1.2  &   4.255  &  1.366  &  1.548  &  1.688  &  2.075      \\
 1.3  &   3.291  &  1.585  &  1.796  &  1.991  &  2.461      \\
 1.4  &   2.605  &  1.834  &  2.079  &  2.296  &  2.857      \\
 1.5  &   2.706  &  2.239  &  2.535  &  2.566  &  3.222      \\
\enddata
\end{deluxetable}

\end{document}